\documentclass{article}

\usepackage[preprint, nonatbib]{neurips_2025}

\usepackage[utf8]{inputenc} % allow utf-8 input
\usepackage[T1]{fontenc}    % use 8-bit T1 fonts
\usepackage{hyperref}       % hyperlinks
\usepackage{url}            % simple URL typesetting
\usepackage{booktabs}       % professional-quality tables
\usepackage{amsfonts}       % blackboard math symbols
\usepackage{nicefrac}       % compact symbols for 1/2, etc.
\usepackage{microtype}      % microtypography
\usepackage{xcolor}         % colors

\usepackage[utf8]{inputenc} % allow utf-8 input
\usepackage[T1]{fontenc}    % use 8-bit T1 fonts
\usepackage{hyperref}       % hyperlinks
\usepackage{url}            % simple URL typesetting
\usepackage{booktabs}       % professional-quality tables
\usepackage{amsfonts}       % blackboard math symbols
\usepackage{nicefrac}       % compact symbols for 1/2, etc.
\usepackage{microtype}      % microtypography
\usepackage{amsmath}
\usepackage{amsthm}
\usepackage{amssymb}
\usepackage{physics}
\usepackage[ruled,vlined]{algorithm2e}
\usepackage{multirow}
\usepackage{graphicx}
\usepackage[belowskip=-6pt]{caption}
\usepackage{graphicx}
\usepackage{amsmath}
\usepackage{subcaption}

\usepackage{graphicx}
\usepackage{amssymb}
\usepackage{multirow}

\makeatletter

\usepackage{xcolor}

\usepackage{hyperref}
\usepackage{url}
\usepackage{graphicx}
\usepackage{xcolor}
\usepackage{amsmath}
\usepackage{amsthm}
\usepackage{amssymb}
\newcommand {\ignore}[1]{}

\usepackage{lineno}
\usepackage{amsmath}
\usepackage{subcaption}

\title{Solving stiff dark matter equations \\ via Jacobian Normalization \\ with Physics-Informed Neural Networks}

\author{M.~P. Bento$^{1, *}$, H.~B. Câmara$^{1,2, \dagger}$, J.~R. Rocha$^{1, *}$, J.~F. Seabra $^{1, *}$ \\
 \\
$^1$Departamento de F\'{\i}sica and CFTP\\
Instituto Superior T\'ecnico, Universidade de Lisboa,
Av. Rovisco Pais 1, 1049-001 Lisboa, Portugal \\ 
$^2$Institute of Experimental and Applied Physics, \\
Czech Technical University in Prague, Prague 160 00, Czech Republic\\
$^*$ \texttt{\{miguel.pedra.bento,jose.r.rocha,joao.f.seabra\}@tecnico.ulisboa.pt}\\
$^\dagger$ \texttt{henrique.camara@cvut.cz}
}

\begin{document}

\maketitle

%%%%%%%%%%%%%%%%%%%%%%%%%%%%%%%%%%%%%%%%%%%%%%%%%%%%%%%%%%%%%%%%%%%%%%%%%%%%%
\begin{abstract}
Stiff differential equations pose a major challenge for Physics-Informed Neural Networks (PINNs), often causing poor convergence. We propose a simple, hyperparameter-free method to address stiffness by normalizing loss residuals with the Jacobian. We provide theoretical indications that \textit{Jacobian-based normalization} can improve gradient descent and validate it on benchmark stiff ordinary differential equations. We then apply it to a realistic system: the stiff Boltzmann equations (BEs) governing weakly interacting massive particle (WIMP) dark matter (DM). Our approach achieves higher accuracy than attention mechanisms previously proposed for handling stiffness, recovering the full solution where prior methods fail. This is further demonstrated in an \textit{inverse} problem with a single experimental data point -- the observed DM relic density -- where our \textit{inverse} PINNs correctly infer the cross section that solves the BEs in both Standard and alternative cosmologies.
\end{abstract}
%%%%%%%%%%%%%%%%%%%%%%%%%%%%%%%%%%%%%%%%%%%%%%%%%%%%%%%%%%%%%%%%%%%%%%%%%%%%%

\maketitle
\noindent

%%%%%%%%%%%%%%%%%%%%%%%%%%%%%%%%%%%%%%%%%%%%%%%%%%%%%%%%%%%%%%%%%%%%%%%%%%%%%
\section{Introduction}
\label{sec:intro}
%%%%%%%%%%%%%%%%%%%%%%%%%%%%%%%%%%%%%%%%%%%%%%%%%%%%%%%%%%%%%%%%%%%%%%%%%%%%%

Physics-Informed Neural Networks (PINNs) exploit the power of neural networks (NNs) as universal function approximators~\cite{HORNIK1989359} by embedding physical knowledge directly into the training process. Their primary goal is to solve ordinary and partial differential equations (ODEs/PDEs)~\cite{raissi2017physicsinformeddeeplearning1,raissi2017physicsinformeddeeplearning2,RAISSI2019686} through the minimization of a physics-informed loss function that combines the residuals of the governing equations with initial, boundary, and final conditions (for a comprehensive review, see~\cite{toscano2024pinns}). This formulation constrains the NN toward physically consistent solutions, improves training efficiency, and reduces dependence on large labeled datasets. A key advantage of PINNs lies in their mesh-free nature, coupled with the use of automatic differentiation (AD)~\cite{baydin2018automaticdifferentiationmachinelearning} to evaluate differential operators with machine precision. PINNs can be applied to both \textit{forward} problems, where the goal is to compute physical solutions from known governing laws, and \textit{inverse} problems, where hidden parameters or constitutive relations are inferred from data. This dual capability has established PINNs as a powerful tool across a wide range of domains: Thermodynamics~\cite{Zobeiry_2021}, Fluid Dynamics~\cite{naderibeni2024learningsolutionsparametricnavierstokes,PhysRevLett.130.244002,wang2025discoveryunstablesingularities}, General Relativity~\cite{Cornell:2022enn,Lobos:2024fzj,Patel:2024iij}, Dark Matter and Cosmology~\cite{Bento:2025agw,Verma:2025ujt}, Astrophysics~\cite{BATY2023100734,BATY2023100734,10.1093/mnras/stad3320,Andika:2024gsc}, etc.

PINNs have also been applied to a variety of stiff problems, including stiff PDEs~\cite{ko2024vspinnfastefficienttraining}, kinetic reaction systems~\cite{Ji_2021,stiff_pinns,10.1063/5.0060697}, and classical benchmarks such as the Van der Pol oscillator~\cite{baty2023solvingstiffordinarydifferential}, among others. Yet stiffness, while a central concept in numerical analysis, remains notoriously difficult to define. Classical perspectives are typically pragmatic. For instance, Ref.~\cite{Hairer} describes stiff problems as those for which explicit schemes fail and implicit solvers are required for stability. In contrast, Ref.~\cite{LambertJ.D1973Cmio} introduces a quantitative measure, the stiffness ratio, defined by the disparities in the magnitudes of the eigenvalues of the Jacobian of the governing equations. Both criteria, however, face limitations. The stiffness ratio becomes ill-defined when the Jacobian is singular, and defining stiffness solely through the breakdown of explicit finite element methods~(FEMs) does not transfer cleanly to PINNs. Nevertheless, empirical evidence clearly shows that stiffness does arise in PINNs, manifesting as characteristic training instabilities~\cite{wang2020understandingmitigatinggradientpathologies,stiff_pinns,Ji_2021,baty2023solvingstiffordinarydifferential}. In practice it often appears as an imbalance in the optimization process. For systems with widely separated time scales, gradients associated with fast-decaying modes dominate backpropagation, suppressing the learning of slower modes and frequently leading to convergence failure~\cite{wang2020understandingmitigatinggradientpathologies,Ji_2021}. To address these issues~\cite{wang2020understandingmitigatinggradientpathologies,basir2022criticalinvestigationfailuremodes}, several extensions of the basic vanilla PINN framework have been proposed. Gradient-balancing methods and adaptive annealing strategies regularize optimization~\cite{baty2023solvingstiffordinarydifferential,ko2024vspinnfastefficienttraining}, while insights from the Neural Tangent Kernel~\cite{WANG2022110768} clarify how stiffness affects the training convergence. Methodological innovations, including attention-based modules such as residual-based~\cite{anagnostopoulos2023residualbasedattentionconnectioninformation} and soft attention~\cite{McClenny_2023}, as well as architecture-level advances such as Transformer-inspired frameworks like PINNsFormer~\cite{zhao2024pinnsformertransformerbasedframeworkphysicsinformed,Wang_2025}, have shown improved ability to capture stiff transients. Moreover, transfer learning has emerged as a complementary strategy, enabling pretrained networks to encode both fast and slow dynamics, thereby accelerating convergence across related parameter stiff regimes~\cite{seiler2025stifftransferlearningphysicsinformed}. These approaches form a growing toolbox for overcoming the unique challenges that stiffness presents in PINNs.

A relevant example of stiff equations arises in particle cosmology, where the evolution of dark matter~(DM) abundance is governed by Boltzmann equations~(BEs) which can be stiff. The Standard Model~(SM) of particle physics fails to account for the observed DM content of the Universe. Astrophysical and cosmological observations show that ordinary matter constitutes only $15\%$ of the total matter, with the remaining $85\%$ being DM~\cite{Planck:2018vyg}. The cold DM (CDM) relic density value is~\cite{Planck:2018vyg}
\begin{equation}
    \Omega_{\text{CDM}} h^2 = 0.120 \pm 0.001 \; .
    \label{eq:Oh2exp}
\end{equation}
In extensions beyond the SM, weakly interacting massive particles (WIMPs) stand out as compelling DM candidates~\cite{Kolb:1988aj,Cirelli:2024ssz}. Initially in thermal equilibrium with the SM plasma, WIMPs undergo thermal freeze-out as the Universe expands and cools. Their relic abundance is set by annihilation processes, $\text{DM} \ \text{DM} \leftrightarrow \text{SM} \ \text{SM}$, characterized by a thermally averaged cross-section $\langle \sigma v \rangle$. The WIMP particle yield evolves according to a stiff, nonlinear BE of Riccati type, which has no known analytical solution. Approximate solutions exist, but accurate predictions for the present-day relic abundance require numerical integration. State-of-the-art tools such as \texttt{MicrOmegas}~\cite{Belanger:2014vza,Belanger:2018ccd} and \texttt{DarkSUSY}~\cite{Bringmann:2018lay} implement FEMs to solve these BEs, incorporating particle physics input to compute observables such as relic density and scattering cross sections. Building on these capabilities, machine learning~(ML) offers a complementary approach to explore the parameter space of DM models~\cite{hepmllivingreview}. While in our previous work~\cite{Bento:2025agw}, we applied PINNs to solve freeze-in DM BEs, which are not inherently stiff, here we focus on the stiff freeze-out problem.

In this work, we address the longstanding challenge of solving stiff differential equations with PINNs. Our contributions can be summarized as follows:
\begin{itemize}
 
\item \textbf{Jacobian-based normalization of residuals:} We introduce a principled normalization of loss residuals based on the Jacobian of the governing differential equation. This approach mitigates stiffness without extra hyperparameters.

\item \textbf{Theoretical and empirical indications of improved convergence:} In Section~\ref{sec:theory} we analyze the Hessian structure of the loss function with and without normalization, finding theoretical indications that Jacobian normalization can improve gradient descent~(GD). Next, in Sections~\ref{sec:simplestode} and~\ref{sec:odes}, we present the empirical results on standard stiff ODE benchmarks, including nonlinear and nonhomogeneous cases, which validate our proposed approach.

\item \textbf{Application to particle cosmology:} In Section~\ref{sec:WIMPDM} we show that Jacobian-normalized PINNs robustly solve stiff WIMP BEs in both Standard and alternative cosmologies, whereas vanilla (Section~\ref{sec:Ablation}) and attention-based (Section~\ref{sec:JacobianVSattention}) PINNs often fail to converge. Our method accurately reproduces DM relic abundance~\eqref{eq:Oh2exp} in the \textit{forward} problem and enables the inference of $\langle \sigma v \rangle$ value in the \textit{inverse} problem as studied in Section~\ref{sec:inverse}.
  
\end{itemize}
Finally, our conclusions are drawn in Section~\ref{sec:concl}.

%%%%%%%%%%%%%%%%%%%%%%%%%%%%%%%%%%%%%%%%%%%%%%%%%%%%%%%%%%%%%%%%%%%%%%%%%%%%%
\section{Jacobian Normalization and Its Effect on PINN Convergence}
\label{sec:theory}
%%%%%%%%%%%%%%%%%%%%%%%%%%%%%%%%%%%%%%%%%%%%%%%%%%%%%%%%%%%%%%%%%%%%%%%%%%%%%

A scalar ODE can be formulated as the problem of finding a function $y$ that satisfies
\begin{equation}
    y: \Omega \to \mathbb{R}, \quad \frac{dy}{dx} = f(y,x), \; y(x_0) = y_0, \quad x \in \Omega \subset \mathbb{R} \; .
    \label{eq:singleODE}
\end{equation}
Our goal is to solve Eq.~\eqref{eq:singleODE} using a PINN  parameterized by $\theta \in \mathbb{R}^n$. Specifically, we aim to approximate its solution through the discretization~\footnote{The elements of the vector $\mathbf{x}$ are commonly referred to as \emph{collocation points}.}
\begin{equation}
    \mathbf{y}(\mathbf{x};\theta) = 
    \begin{bmatrix} y(x_1;\theta) \\ \vdots \\ y(x_m;\theta) \end{bmatrix} \in \mathbb{R}^{m}, \quad 
    \mathbf{x} = \begin{bmatrix} x_1 \\ \vdots \\ x_m \end{bmatrix} \in \Omega^m \subset \mathbb{R}^m \; .
\end{equation}
The parameters $\theta$ are obtained by minimizing the loss function,
\begin{equation}
    \mathcal{L} = \mathcal{L}_{\mathrm{r}} + \mathcal{L}_{\mathrm{ic}} \, ,
    \label{eq:totalloss}
\end{equation}
where the first term corresponds to the mean squared error (MSE) of the ODE residual,  
\begin{equation}
    \mathcal{L}_{\mathrm{r}} = \frac{1}{m} \sum_{k=1}^m R_k^2[y(x_k;\theta),x_k] = \frac{1}{m}\sum_{k=1}^m 
    \left\{ \frac{\partial y}{\partial x}(x_k;\theta) - f[y(x_k,\theta),x_k] \right\}^2 \, ,
    \label{eq:lossresiduals}
\end{equation}
with $R \in \mathbb{R}^m$ denoting the residual vector.  
The second term,  
\begin{equation}
    \mathcal{L}_{\mathrm{ic}} = \left[y(x_0) - y_0 \right]^2 \, ,
    \label{eq:lossinitialcondition}
\end{equation}
ensures that the initial condition is satisfied. However, when the ODE~\eqref{eq:singleODE} is stiff, the optimization becomes unstable and convergence cannot be achieved. Stiffness in an ODE is known to be governed by the magnitude of its Jacobian~\cite{LambertJ.D1973Cmio,Ji_2021}, defined as  
\begin{align}
    J_y := \frac{df}{dy} \; .
\end{align}
In particular, stiffness arises when $J_y \ll 0$. While the role of $J_y$ in classical numerical methods is well understood, appearing through step-size constraints, its impact in PINNs is less clear. Nonetheless, we find that the two are closely related: when attempting to solve the simple ODE
\begin{align}
\frac{dy}{dx} &= -Cy \; , \quad y(0)=1 \; , \label{eq:case1}
\end{align}
with a PINN, large negative values of $J_y = -C$ cause the training to fail to converge to the correct solution (see Fig.~\ref{fig:1stToy}). One way to interpret this issue is to note that, in this simple case, stiffness is determined by a single constant $C$. 
Its magnitude directly governs the scaling of the eigenvalues $\lambda_i$ of the Hessian,
\begin{equation}
    H_{ij} = \frac{\partial^2 \mathcal{L}_{\mathrm{r}}}{\partial \theta_i \, \partial \theta_j} \; , \quad H \in \mathbb{R}^{n \times n} \; ,
    \label{eq:HessianGeneric}
\end{equation}
as we will show below. Consequently, as $J_y$ increases in magnitude, the corresponding eigenvalues~$\lambda_i$, evaluated at the loss minima, also become larger. A smaller learning rate $\eta$ is therefore required for convergence in vanilla GD, 
according to the bound~\cite{bishop2023learning}
\begin{equation}
    |1 - \eta \lambda_i| < 1 \; .
    \label{eq:BoundBishop}
\end{equation}
However, it is always possible to apply a normalization, analogous to the input/output scaling commonly employed in ML, which can be defined as  
\begin{equation}
    R \rightarrow \frac{R}{C} 
    \Rightarrow 
    \lambda_i \rightarrow \frac{\lambda_i}{C^2}
    .
    \label{eq:NewNormalization}
\end{equation}
This normalization effectively divides the eigenvalues $\lambda_i$ by $C^2$, or equivalently, according to Eq.~\eqref{eq:BoundBishop}, scales the learning rate by $C^{-2}$, thus improving convergence by a considerable factor. Although the analysis we presented was derived for vanilla GD, it also applies to more advanced optimization algorithms as shown in Ref.~\cite{wang2020understandingmitigatinggradientpathologies}.

To make the previous argument precise, we compute the Hessian of the residual loss. Replacing Eq.~\eqref{eq:lossresiduals} into the generic definition of the Hessian~\eqref{eq:HessianGeneric}, we obtain
\begin{align}
\label{eq:hessian_residual}
    (H_{\mathrm{r}})_{ij} = G_{ij} + K_{ij} \, ,
\end{align}
where
\begin{align}
\label{eq:hessian_GK}
    G_{ij} &= \frac{2}{m} \sum_{k=1}^m 
    \left[
        \frac{d}{dx}\!\left(\frac{\partial y_k}{\partial \theta_i}\right)
        - J_y \frac{\partial y_k}{\partial \theta_i}
    \right]
    \left[
        \frac{d}{dx}\!\left(\frac{\partial y_k}{\partial \theta_j}\right)
        - J_y \frac{\partial y_k}{\partial \theta_j}
    \right] ,
    \\
    K_{ij} &= \frac{2}{m} \sum_{k=1}^m R_k
    \left[
        \frac{d}{dx}\!\left(\frac{\partial^2 y_k}{\partial \theta_i \partial \theta_j}\right)
        - J_y \frac{\partial^2 y_k}{\partial \theta_i \partial \theta_j} 
        - \frac{d J_y}{d y} \frac{\partial y_k}{\partial \theta_i } \frac{\partial y_k}{\partial \theta_j}
    \right] ,
\end{align}
with $y_k \equiv y(x_k,\theta)$. Here $\mathbf{G} \in \mathbb{R}^{n \times n}$ corresponds to the Gauss Newton term, and $\mathbf{K} \in \mathbb{R}^{n \times n}$ is proportional to the residual and therefore becomes small near a minimum. Moreover, $\mathbf{G}$ contains an explicit factor proportional to $J_y^2$. However, additional implicit dependences on $J_y$ may arise through other factors, but we expect them to contribute with smaller powers. We therefore write the residual Hessian scaling schematically as
\begin{align}
    (H_{\mathrm{r}})_{ij} = \Theta\!\left(J_y^{\,2 + a_{ij}}\right) ,
\end{align}
where $a_{ij}$ parametrizes possible additional contributions to the scaling. Here $\Theta(\cdot)$ denotes asymptotic scaling for large $|J_y|$.
Taking $a = \max(a_{ij})$, the Frobenius norm satisfies
\begin{align}
    \norm{H_{\mathrm{r}}}_F = \Theta \left(\left|J_y^{2+a}\right|\right) .
    \label{eq:Asymptotic}
\end{align}
Taking into consideration that $H_r$ is a real symmetric matrix, we find that
\begin{align}
 \norm{H_{\mathrm{r}}}_F^2
    =  \sum_{i=1}^n \lambda_i^2(H_r)
    \Rightarrow
    \lambda_{\max}^2(H_{\mathrm{r}}) 
    \le \norm{H_{\mathrm{r}}}_F^2
    \le n \lambda_{\max}^2(H_{\mathrm{r}}) \;,
\end{align}
or equivalently
\begin{align}
    \frac{1}{\sqrt{n}}\, \norm{H_{\mathrm{r}}}_F
    \le \left|\lambda_{\max}(H_{\mathrm{r}})\right|
    \le \norm{H_{\mathrm{r}}}_F \,.
\end{align}
Consequently, considering Eq.~\eqref{eq:Asymptotic}, we find that for the asymptotic behavior
\begin{align}
    |\lambda_{\max}(H_{\mathrm{r}})| = \Theta\left(\left|J_y^{2+a}\right|\right) .
    \label{eq:JyPower}
\end{align}
This means that, in the stiff regime where $|J_y|$ is large, $\lambda_{\max}(H_{\mathrm{r}})$ increases accordingly.

Up to this point we have focused on the residual part of the loss and have not included the contribution from the initial condition. The full Hessian of the total loss can be written as
\begin{align}
    H = H_{\mathrm{r}} + H_{\mathrm{ic}} \, ,
\end{align}
where $H_{\mathrm{ic}}$ is the Hessian of $\mathcal{L}_{\mathrm{ic}}$. Following Weyl's inequality for sums of symmetric matrices, one obtains
\begin{align}
    \lambda_{\min}(H_{\mathrm{ic}}) + \lambda_{\max}(H_{\mathrm{r}})
    \le \lambda_{\max}(H)
    \le \lambda_{\max}(H_{\mathrm{ic}}) + \lambda_{\max}(H_{\mathrm{r}}) \, .
    \label{eq:Weyl}
\end{align}
In the stiff regime, $\lambda_{\max}(H_{\mathrm{r}})$ grows with the Jacobian magnitude [see Eq.~\eqref{eq:JyPower}]. Since $H_{\mathrm{ic}}$ does not contain the Jacobian factor present in $H_{\mathrm{r}}$, the Hessian of the total loss is dominated by the residual term, so it follows from Eq.~\eqref{eq:Weyl} that
\begin{align}
    \lambda_{\max}(H) = \Theta(\left|J_y^{2+a}\right|) \, .
    \label{eq:lamaxfullHessian}
\end{align}
Given the stability condition in Eq.~\eqref{eq:BoundBishop}, an increase in $\lambda_{\max}(H)$ requires a smaller learning rate $\eta$. However, if $\eta$ is chosen sufficiently small to compensate for the magnitude of $\lambda_{\max}(H)$, convergence becomes very slow. Conversely, if $\eta$ exceeds the stability bound, GD updates fail to converge.

Another perspective on the convergence issue for large Jacobian follows from inspecting the gradient of the residual loss, given by
\begin{align}
    \frac{\partial \mathcal L_{\mathrm r}}{\partial \theta_i}
    =
    \frac{2}{m}\sum_{k=1}^m
    R_k
    \left[
        \frac{d}{dx}\!\left(\frac{\partial y_k}{\partial \theta_i}\right)
        - J_y \frac{\partial y_k}{\partial \theta_i}
    \right].
\end{align}
In the early and intermediate stages of training, when the residual $R_k$ is not yet small, a large value of $|J_y|$ amplifies the residual loss gradient. As $|J_y|$ increases, one typically observes that
\begin{align}
    \left|\frac{\partial \mathcal L_{\mathrm r}}{\partial \theta_i}\right|
    \gg
    \left|\frac{\partial \mathcal L_{\mathrm{ic}}}{\partial \theta_i}\right| ,
\end{align}
so the parameter updates are dominated by the residual term. Since $\mathcal L_{\mathrm r}$ is accumulated over all collocation points while $\mathcal L_{\mathrm{ic}}$ is imposed at a single point, training tends to focus on reducing $\mathcal L_{\mathrm r}$ throughout the interval, and the initial condition is enforced more slowly.

Motivated by this analysis, we normalize the residual contribution to the loss as
\begin{equation}
    \mathcal{L}_{\mathrm{r}} :=
    \frac{1}{m}\sum_{k=1}^m 
    \frac{\left[ \frac{\partial y_k}{\partial x} - f(y_k,x_k) \right]^2}
         {1 + J_y^2} ,
    \label{eq:lossresidualsnormalization}
\end{equation}
where the denominator ensures that $\mathcal{L}_{\mathrm{r}}$ remains finite and reduces to its unnormalized form~\eqref{eq:lossresiduals} in the limit of a vanishing Jacobian. This rescaling suppresses the $J_y^2$ growth of the residual Hessian eigenvalues~\eqref{eq:JyPower}, mitigating stiffness. Additionally, the normalization reduces the imbalance between the residual and initial condition contributions during training.

%%%%%%%%%%%%%%%%%%%%%%%%%%%%%%%%%%%%%%%%%%%%%%%%%%%%%%%%%%%%%%%%%%%%%%%%%%%%%
\subsection{A benchmark on the simplest class of stiff ODEs}
\label{sec:simplestode}
%%%%%%%%%%%%%%%%%%%%%%%%%%%%%%%%%%%%%%%%%%%%%%%%%%%%%%%%%%%%%%%%%%%%%%%%%%%%%

In this section, we train PINNs on the simplest class of ODEs~\eqref{eq:case1}. The vanilla (unnormalized) PINN minimizes the total loss function defined in Eqs.~\eqref{eq:totalloss}, \eqref{eq:lossresiduals}, and~\eqref{eq:lossinitialcondition}. In contrast, for the Jacobian-normalized PINN, the residual component of the loss is given by Eq.~\eqref{eq:lossresidualsnormalization} with $f(y,x)=-C y$ and $J_y = - C$. To benchmark the performance of the PINNs we compute the mean squared difference
\begin{equation}
\overline{\varepsilon} = \frac{1}{m}\sum_{k=1}^m \left[y(x_k)^{\text{PINN}} - y(x_k)^{\text{FEM}} \right]^2  \; ,
\label{eq:epsilonMSEToy}
\end{equation}
between PINN and FEM solutions, obtained using the initial-value ODE solver \texttt{solve\_ivp} from the \texttt{SciPy} library, employing the implicit backward differentiation formula (BDF) method\footnote{The benchmark stiff ODEs in Eqs.~\eqref{eq:case1},~\eqref{eq:case2} and~\eqref{eq:case3} possess closed-form analytical solutions, which agree with the numerical results produced by the BDF implicit method.}.

The vanilla and Jacobian-normalized PINNs are trained for twenty values of $C$, logarithmically spaced within $[1, 10^4]$. The architecture consists of four hidden layers with 40 units each, using Gaussian Error Linear Unit (GELU) activation functions. Training minimizes the total loss function evaluated at $m=1000$ collocation points logarithmically sampled in $x \in [e^{-15},1]$. For optimization, up to $5 \times 10^5$ epochs are performed, with early stopping triggered once the total loss falls below $10^{-8}$. Adam optimizer is used with an exponentially decaying learning rate, initialized at $\eta =8\times 10^{-5}$ and reduced by a factor of $0.99$ every 1000 epochs.

    \begin{figure*}[t!]
        \centering
        \includegraphics[scale=0.29]{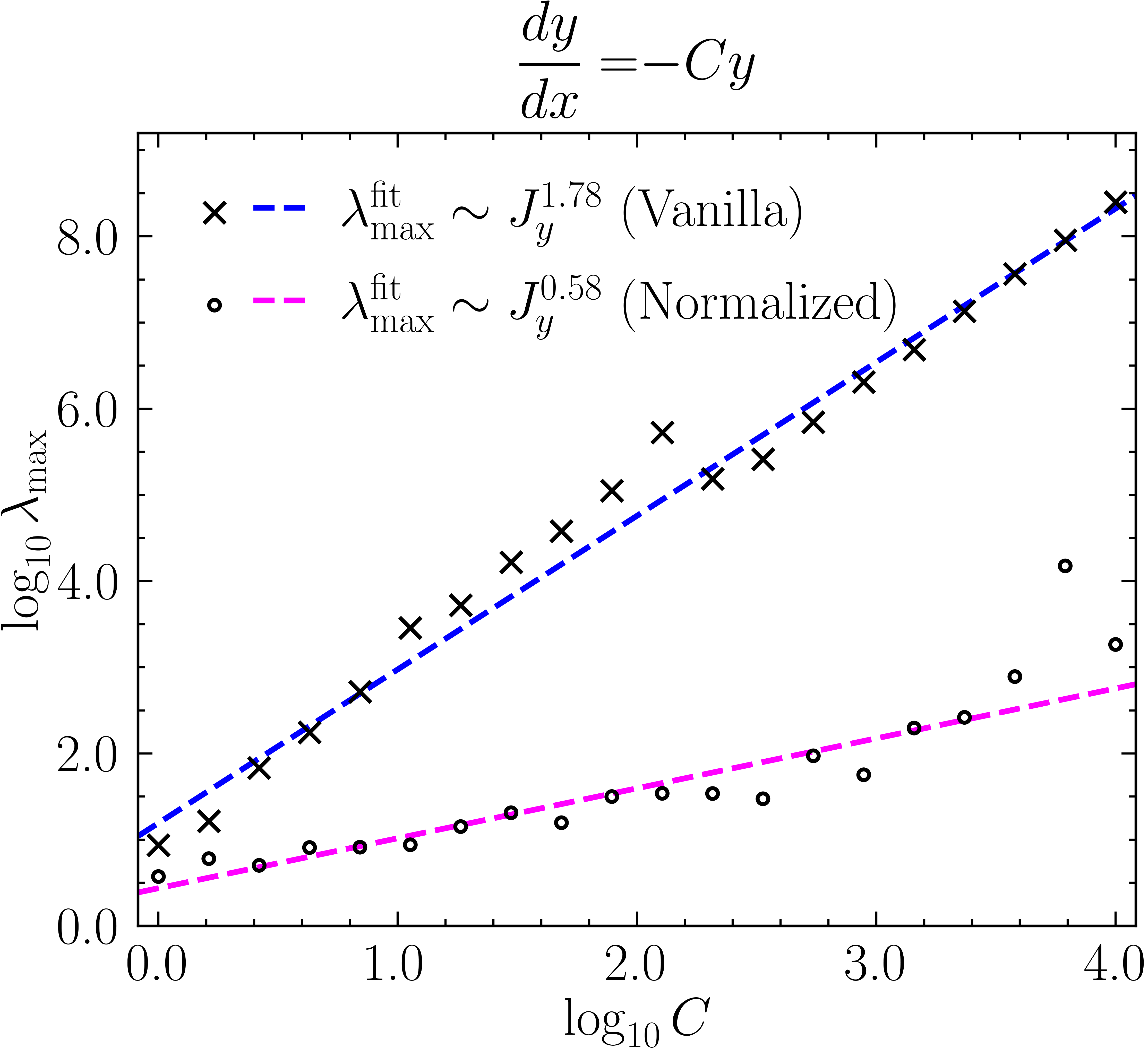} \hspace{+0.1cm}\includegraphics[scale=0.29]{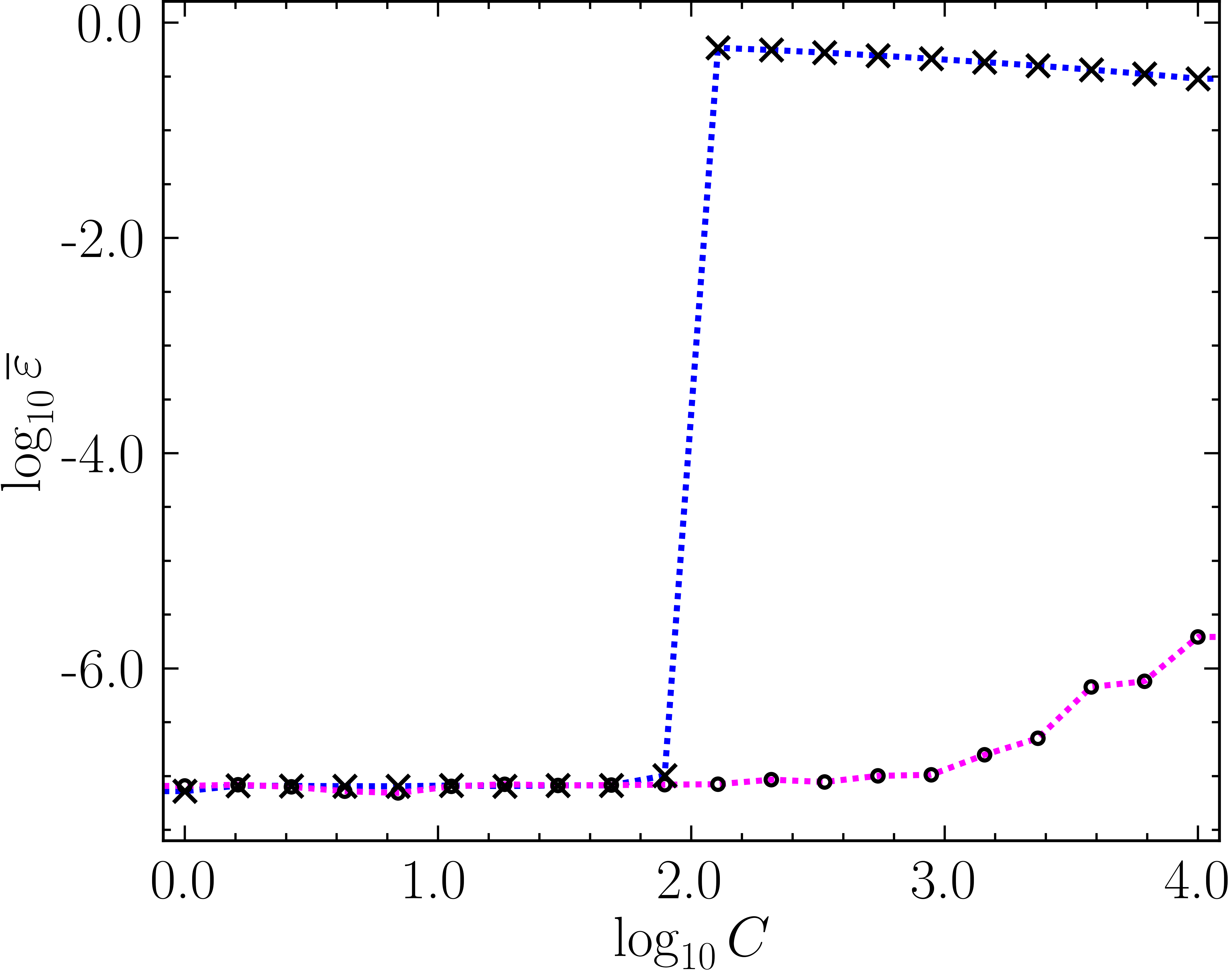}
        \caption{PINN results for ODE~\eqref{eq:case1}, with normalized (unnormalized) residuals in green (red). Left: Maximum Hessian eigenvalue $\lambda_{\text{max}}$~\eqref{eq:HessianGeneric} in terms of the Jacobian $C$. The circles (crosses) indicate normalized (unnormalized) residuals. The linear fit slope gives the power-law scaling exponent of $\lambda_{\text{max}}$ with $C$. Right: Mean squared difference between PINN and FEM solutions~\eqref{eq:epsilonMSEToy} in terms of $C$.}
    \label{fig:1stToy}
    \end{figure*}
In Fig.~\ref{fig:1stToy} we show the results of training for the normalized (unnormalized) PINN in magenta (blue). In the left panel, we show the maximum eigenvalue of the Hessian $\lambda_{\text{max}}$~\eqref{eq:HessianGeneric}, computed via the power iteration algorithm, as a function of the absolute value of the Jacobian $|J_y|$ (which for Eq.~\eqref{eq:case1} equals $C$). To quantify the dependence of $\lambda_{\text{max}}$ on $C$, the data are fitted using a Huber regressor, chosen for its robustness to outliers. Without normalization, the fit to the data yields $\lambda_{\text{max}} \sim \mathcal{O}(J_y^{1.78})$, showing that $\lambda_{\text{max}}$ scales approximately as $C^2$, whereas for the normalized residual loss this scaling is considerably weaker $\lambda_{\text{max}} \sim \mathcal{O}(J_y^{0.58})$. This result is consistent with the theoretical behavior of the Hessian shown in Eqs.~\eqref{eq:JyPower} and~\eqref{eq:lamaxfullHessian}. The right panel shows $\overline{\varepsilon}$~\eqref{eq:epsilonMSEToy} in terms of~$C$. We notice a discrepancy between the normalized and vanilla cases starting at $C \approx 10^{2}$. While the normalized PINN maintains $\overline{\varepsilon} \approx 10^{-7}$, the unnormalized PINN error increases by six orders of magnitude to $\overline{\varepsilon} \approx 10^{-0.5}$. This breakdown with increasing stiffness is consistent with Eq.~\eqref{eq:BoundBishop}, which predicts that convergence may deteriorate or fail once Hessian eigenvalues become sufficiently large. For the largest $C$ values, the vanilla network collapses to the trivial solution $y(x_k)=0$, minimizing residuals while violating the initial condition. In contrast, the normalized PINN accurately reproduces the FEM solutions for all tested $C$, demonstrating effective stiffness mitigation.

%%%%%%%%%%%%%%%%%%%%%%%%%%%%%%%%%%%%%%%%%%%%%%%%%%%%%%%%%%%%%%%%%%%%%%%%%%%%%
\subsection{Benchmarks on nonhomogeneous and nonlinear ODEs}
\label{sec:odes}
%%%%%%%%%%%%%%%%%%%%%%%%%%%%%%%%%%%%%%%%%%%%%%%%%%%%%%%%%%%%%%%%%%%%%%%%%%%%%

%
    \begin{figure*}[t!]
        \centering
        \includegraphics[scale=0.29]{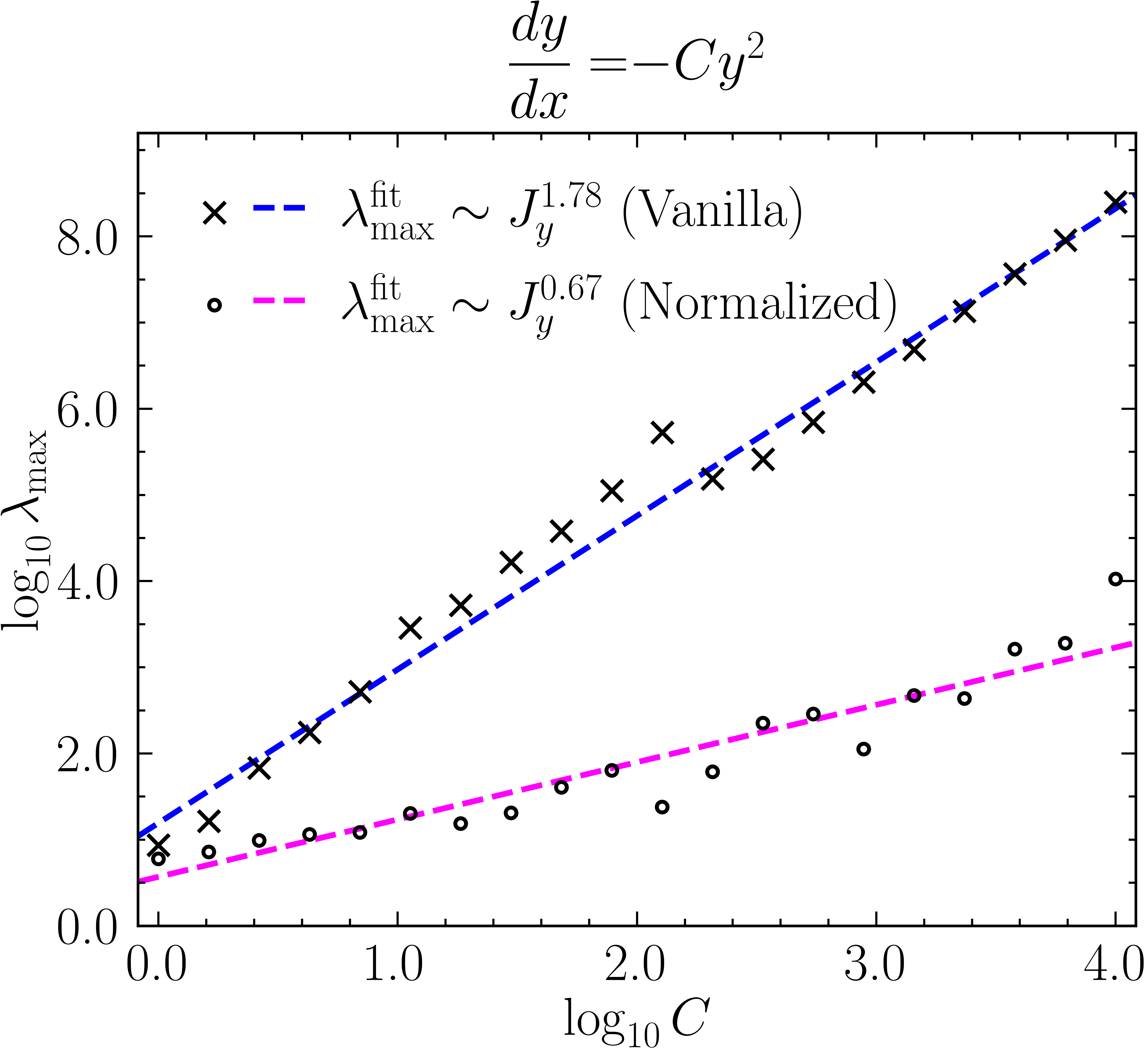} \hspace{+0.1cm} \includegraphics[scale=0.29]{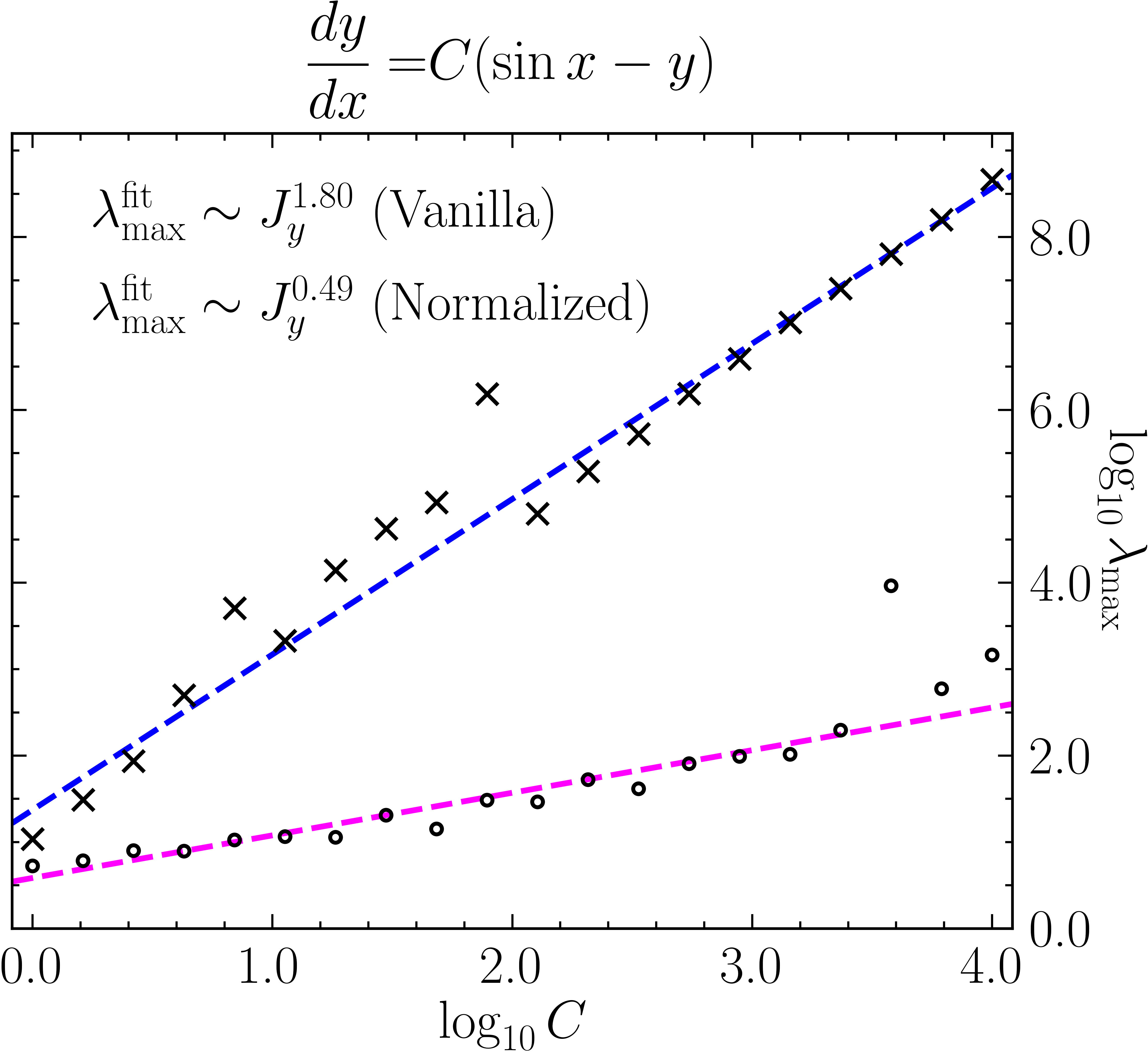} \\
        \vspace{+0.2cm} \includegraphics[scale=0.29]{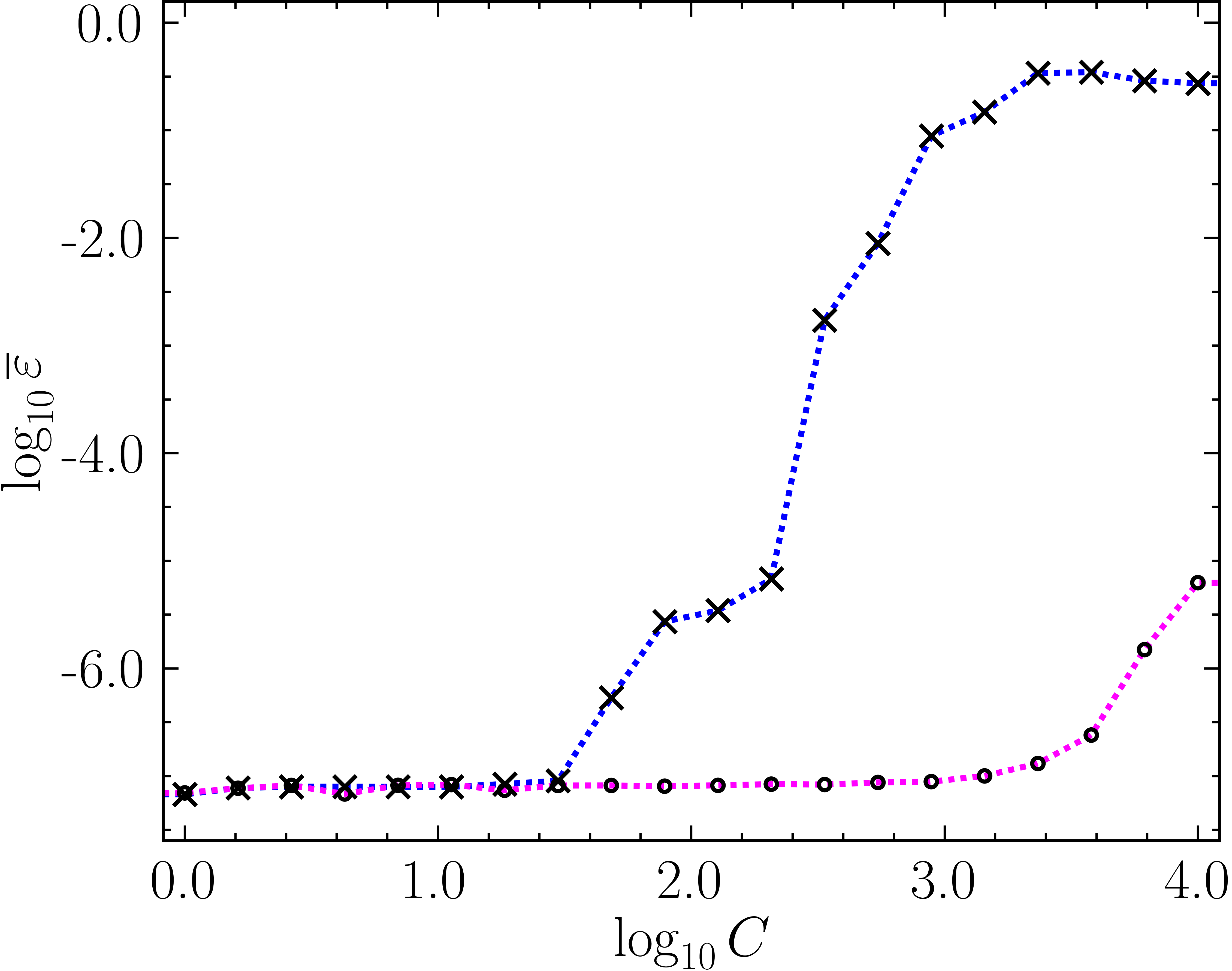}  \hspace{+0.09cm}\includegraphics[scale=0.29]{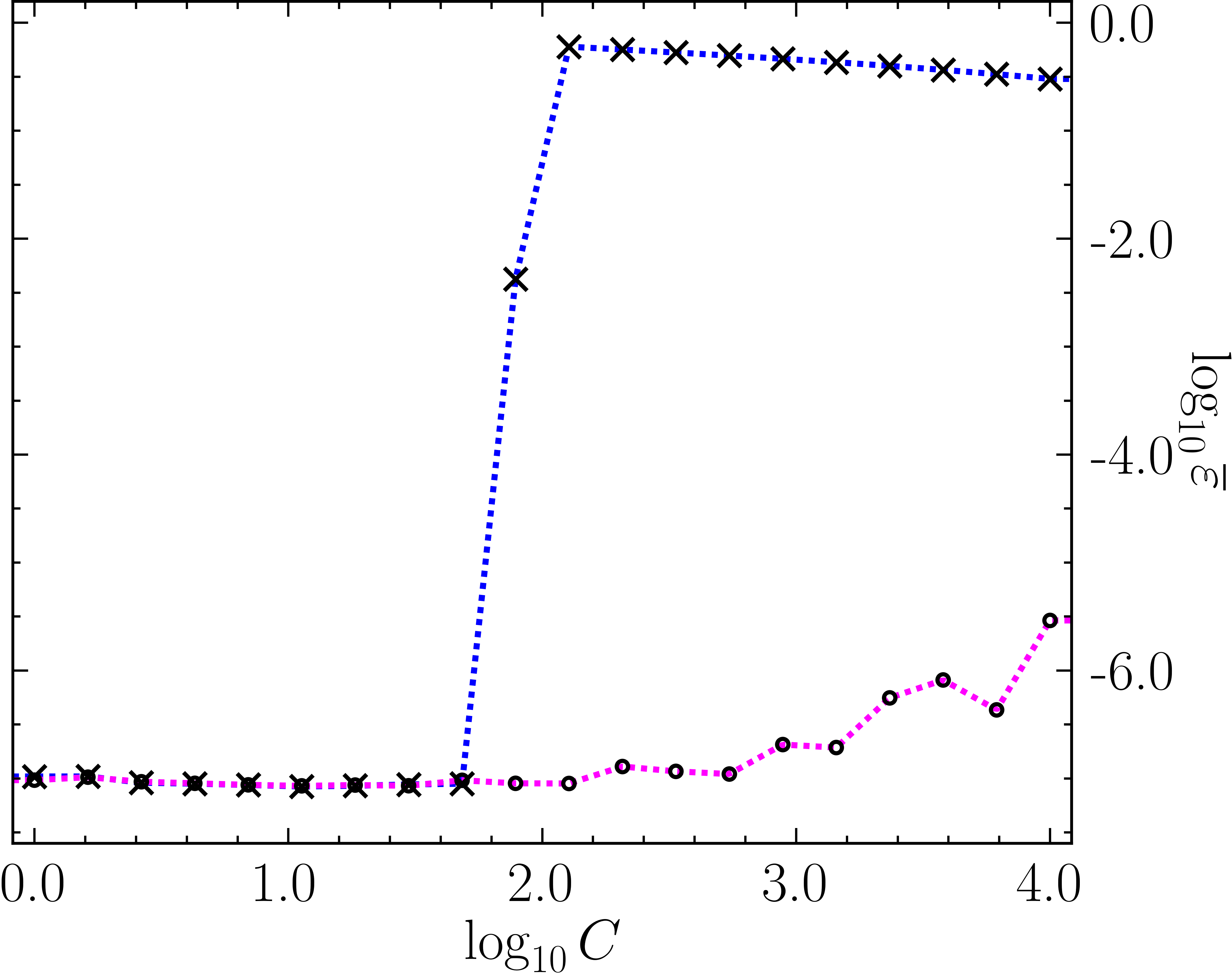} 
        \caption{PINN results for ODEs Eq.~\eqref{eq:case2} (left) and~\eqref{eq:case3} (right), with normalized [unnormalized] residuals in magenta [blue]. Top panels: Maximum Hessian eigenvalue $\lambda_{\text{max}}$~\eqref{eq:HessianGeneric} in terms of $C$. Bottom panels: Mean squared difference between PINN and FEM solutions $\overline{\varepsilon}$~\eqref{eq:epsilonMSEToy} in terms of $C$.}
    \label{fig:1storderODEs}
    \end{figure*}
We continue our empirical study of Jacobian normalization and stiffness by considering the following nonlinear, nonhomogeneous first-order ODEs:
\begin{alignat}{3}
\frac{dy}{dx} &= -C y^2\; ,        &\quad y(0)&=1\; , &\quad J_y &= -2C y \; , \label{eq:case2}\\
\frac{dy}{dx} &= C(\sin x - y)\; , &\quad y(0)&=1\; , &\quad J_y &= -C\; ,    \label{eq:case3}
\end{alignat}
where increasing $C$ also increases the magnitude of $J_y$, thereby leading to a stiffer regime. The PINNs settings for solving the above ODEs are the same as those described in Section~\ref{sec:simplestode}. 

In Fig.~\ref{fig:1storderODEs} we present the results of PINNs trained to solve the ODE~\eqref{eq:case2} (left) and~\eqref{eq:case3} (right), for both Jacobian-normalized (magenta) and vanilla (blue) cases. The top panels show $\lambda_{\text{max}}$ as a function of the $J_y$ which is proportional to $C$ for both ODEs. The vanilla PINNs exhibits $\lambda_{\text{max}} \sim \mathcal{O}(J_y^{1.8\text{–}1.9})$ scaling, whereas normalization reduces the exponent to $\lambda_{\text{max}} \sim \mathcal{O}(J_y^{0.5\text{–}0.7})$, in agreement with the theoretical expectations of Eqs.~\eqref{eq:JyPower} and~\eqref{eq:lamaxfullHessian}. As discussed previously, this implies that the normalized loss converges more readily to the correct solution, in accordance with Eq.~\eqref{eq:BoundBishop}. This behavior is further confirmed in the bottom panels, which show $\overline{\varepsilon}$~\eqref{eq:epsilonMSEToy} as a function of $C$. Namely, in the nonlinear ODE case (bottom left panel), $\overline{\varepsilon}$ for the unnormalized PINN begins to increase at $C \approx 10^{2}$, reaching $\overline{\varepsilon} \approx 10^{-0.5}$ at $C \approx 10^{3}$. A similar behavior is observed for the nonhomogeneous ODE (bottom right panel), for the vanilla case $\overline{\varepsilon}$ steeply reaches the value $\overline{\varepsilon} \approx 10^{-0.5}$ from $C \approx 10^{2}$, where the solution enters a regime in which the initial condition is no longer satisfied. In contrast, the normalized PINN maintains a mean squared difference between the PINN and FEM solutions of order $10^{-7}$ across the entire range of tested Jacobian values, increasing only to $\mathcal{O}(10^{-5})$ for $C \gtrsim 10^{3.5}$ in the nonlinear case. The behavior of the PINNs for both ODEs closely mirrors that observed in the previous section for the ODE~\eqref{eq:case1}.

Overall, the empirical study conducted in Sections~\ref{sec:simplestode} and~\ref{sec:odes}, across a set of benchmark stiff problems, validates \textit{Jacobian-based normalization of the residual loss} as a principled and straightforward strategy for handling stiff equations.

%%%%%%%%%%%%%%%%%%%%%%%%%%%%%%%%%%%%%%%%%%%%%%%%%%%%%%%%%%%%%%%%%%%%%%%%%%%%%
\section{A Concrete Application: WIMP Dark Matter in Alternative Cosmologies}
\label{sec:WIMPDM}
%%%%%%%%%%%%%%%%%%%%%%%%%%%%%%%%%%%%%%%%%%%%%%%%%%%%%%%%%%%%%%%%%%%%%%%%%%%%%

In this section we apply \emph{Jacobian-based normalization of residuals} to solve a realistic theoretical physics problem with PINNs: the paradigmatic freeze-out BE for WIMP DM in Standard and alternative cosmological scenarios.

The WIMP particle yield $Y$ evolves according to the BE:
\begin{equation}
    \frac{dY(x)}{dx} = - \frac{1}{x^2} \frac{\langle \sigma v \rangle S(M)}{H(M)} \left[Y(x)^2 - Y_{\text{eq}}(x)^2\right] \; , \; Y_0 \equiv Y(x_0) = Y_{\text{eq}}(x_0) \; ,
\label{eq:freezeout1}
\end{equation}
where $x \equiv M/T$, with $M$ being the mass of the DM candidate and $T$ the temperature. For our purposes, we take the thermally averaged cross-section $\langle \sigma v \rangle$ to be simply a positive constant. The initial condition $Y_0$ is defined at an initial early time $x_0$, via the equilibrium yield $Y_{\text{eq}}(x)$, which in the non-relativistic limit $x \gg 3$ is given by
\begin{equation}
    Y_{\text{eq}}(x) = \frac{45}{4 \pi^4} \frac{g}{g_{\ast S}} x^2 \mathcal{K}_2(x) \xrightarrow[x \gg 3]{} \frac{45}{\sqrt{32 \pi^7}} \frac{g}{g_{\ast S}} x^{3/2} \exp(-x) \; .
    \label{eq:Yeq}
\end{equation}
Here, $\mathcal{K}_n$ is the modified Bessel function of second kind of order $n$, while\footnote{We consider, without loss of generality, a scalar DM candidate, hence we set $g=1$.} $g=1$ and $g_{\ast S} \simeq 100$ correspond to the number of internal and effective degrees of freedom, respectively. 

In Standard cosmology, the Hubble expansion rate $H(T)$ and entropy density $S(T)$ are
\begin{align}
    H(T) = \sqrt{\frac{4\pi^3}{45}} \, g_\ast^{1/2} \frac{T^2}{M_{\text{Pl}}} \; , \quad S(T) = \frac{2\pi^2}{45} \, g_{\ast S} T^3 \; ,
\label{eq:Hands}
\end{align}
where $g_\ast \simeq g_{\ast S} \simeq 100$ and $M_{\text{Pl}} = 1.22 \times 10^{19}~\mathrm{GeV}$ is the Planck mass. From a theoretical perspective, it is interesting to understand what happens when the Standard law of cosmology is modified. To capture such deviations we consider an alternative expansion history parameterized via a switch-like function~\cite{Okada:2004nc,Okada:2009xe,Liu:2023qhv}:
\begin{align}
    H(x) \rightarrow H(x) \times F(x,x_t,\gamma) \; , \;
    F(x,x_t,\gamma) = \left\{  
    \begin{array}{ll}  \displaystyle
    \left(x_t/x\right)^\gamma & \ \text{for} \ x < x_t \vspace{0.3cm} \\
    \displaystyle \hspace{0.3cm} 1 \hspace{0.5cm} & \ \text{for} \ x > x_t
    \end{array} \right. \; ,
\label{eq:Hpowerlaw}
\end{align}
introducing a transition temperature $T_t = M/x_t$ and a power-law index $\gamma$. This approximation models early Universe modifications arising in extra-dimensional string theory inspired scenarios, e.g., Gauss-Bonnet~(GB) ($\gamma = -2/3$) and Randall-Sundrum~(RS) ($\gamma = 2$) braneworld cosmologies~\cite{Okada:2004nc,Okada:2009xe,Liu:2023qhv}.

Solving Eq.~\eqref{eq:freezeout1} yields the DM abundance $Y(x)$, from which the present-day relic density is:
\begin{equation}
    \Omega h^2 = \frac{S_0}{\rho_0/h^2} M Y(x \rightarrow \infty) \; ,
    \label{eq:relic}
\end{equation}
with $S_0 = 2.89 \times 10^3~\text{cm}^{-3}$ and $\rho_0 = 1.05 h^2 \times 10^{-5}~\text{GeV}/\text{cm}^{3}$ being the present entropy and critical energy density, respectively~\cite{ParticleDataGroup:2024cfk}. Eq.~\eqref{eq:freezeout1} represents a specific form of the Riccati equation, which for WIMP DM is a stiff system with no known analytical solution. However, one can obtain the instructive approximate solution for the final relic abundance~\cite{Kolb:1990vq}:
\begin{equation}
  \Omega h^2 \simeq \frac{S_0}{\rho_0/h^2} \sqrt{\frac{45}{\pi}} \frac{g_\ast^{1/2}}{g_{\ast S}} \frac{x_d}{M_{\text{Pl}}\langle \sigma v \rangle} \simeq 0.12 \left( \frac{x_d}{25} \right) \left(\frac{1.78 \times 10^{-9} \ \text{GeV}^{-2}}{\langle \sigma v \rangle}\right) \; ,
  \label{eq:analyticalapprox}
\end{equation}
where $x_d \equiv M/T_d$ is the freeze-out temperature that characterizes the time at which $Y$ ceases to track the equilibrium $Y_{\text{eq}}$. The value of $\langle \sigma v \rangle$ reproducing the CDM relic data point of Eq.~\eqref{eq:Oh2exp} is around $\mathcal{O}(10^{-9})$ GeV$^{-2}$ (see Section~\ref{sec:inverse}) or equivalently $\mathcal{O}(1)$~pb, a typical weak scale scattering cross section value. Considering a DM candidate of mass $M=100$ GeV, the resulting Jacobian 
\begin{equation}
    J_y \propto - M_{\text{Pl}} \ M \ \langle \sigma v \rangle \sim - \mathcal{O}(10^{12}) \; ,
    \label{eq:largejacobian}
\end{equation}
takes a large negative value. Thus, in order to solve the freeze-out BE with PINNs it is necessary to apply stiffness strategies, among which is our proposal of \emph{Jacobian-based normalization of residuals}.

%%%%%%%%%%%%%%%%%%%%%%%%%%%%%%%%%%%%%%%%%%%%%%%%%%%%%%%%%%%%%%%%%%%%%%%%%%%%%
\subsection{Recasting the freeze-out Boltzmann equation}
\label{sec:Recast}
%%%%%%%%%%%%%%%%%%%%%%%%%%%%%%%%%%%%%%%%%%%%%%%%%%%%%%%%%%%%%%%%%%%%%%%%%%%%%

Our study of the WIMP DM paradigm in Standard and alternative cosmologies requires solving the BE~\eqref{eq:freezeout1} taking into account Eq.~\eqref{eq:Hpowerlaw}. This is done within the interval $1\leq x\leq 5\times 10^4$ for $M=100$ GeV. Using Eq.~\eqref{eq:Yeq} we obtain the initial condition $Y_0 = Y_{\text{eq}}(x_0 = 1)=5.32\times 10^{-4}$. Furthermore, from the current measurement of the relic density~\eqref{eq:Oh2exp} and using Eq.~\eqref{eq:relic}, we compute $Y(x\rightarrow \infty)_{\Omega h^2} = 4.36 \times 10^{-12}$. Due to the broad range of values for $x$ and $Y$, spanning over multiple orders of magnitude, we introduce the following change of variables: $z \equiv \ln x$ and $W(z) \equiv \ln Y(z)$. Further improvements to the performance of the PINNs can be achieved by normalizing $W_n(z) = -W(z)/W_{\Omega h^2}$, with $W_{\Omega h^2} \equiv \ln Y(x \rightarrow \infty)$. We also define:
\begin{align}
    \exp[C(M, \langle \sigma v \rangle)] \equiv\frac{\langle \sigma v \rangle S(M)}{H(M)} \; ,
    \label{eq:Cdef}
\end{align}
which encodes the particle interaction strength via $\langle \sigma v \rangle$, and the function [see Eq.~\eqref{eq:Hpowerlaw}]
\begin{equation}
    F(z, z_t, \gamma) = \exp\left[\gamma \ \text{ReLU}(z_t-z)\right] \; ,
    \label{eq:Fnew}
\end{equation}
with "ReLU" being the Rectified Linear Unit and $z_t = \ln x_t $.

The BE for WIMP DM in alternative cosmologies can now be written as:
\begin{align}
\mathcal{E}\left[z, W, \frac{d W}{dz} ; C, z_t, \gamma\right] & = 0 \;  \nonumber \\
\Leftrightarrow \frac{d W}{dz} + \exp\left[C - \gamma \ \text{ReLU}(z_t-z) - z\right] \times \left[\exp(W) - \exp(2 W_{\text{eq}} - W) \right] & = 0 \; ,
\label{eq:freezeout2}
\end{align}
where by setting $\gamma \to 0$ we recover the Standard cosmological scenario. 

%%%%%%%%%%%%%%%%%%%%%%%%%%%%%%%%%%%%%%%%%%%%%%%%%%%%%%%%%%%%%%%%%%%%%%%%%%%%%
\subsection{Ablation test: Jacobian normalization versus vanilla PINN}
\label{sec:Ablation}
%%%%%%%%%%%%%%%%%%%%%%%%%%%%%%%%%%%%%%%%%%%%%%%%%%%%%%%%%%%%%%%%%%%%%%%%%%%%%

%
\begin{figure*}[t!]
    \centering
    \includegraphics[scale=0.29]{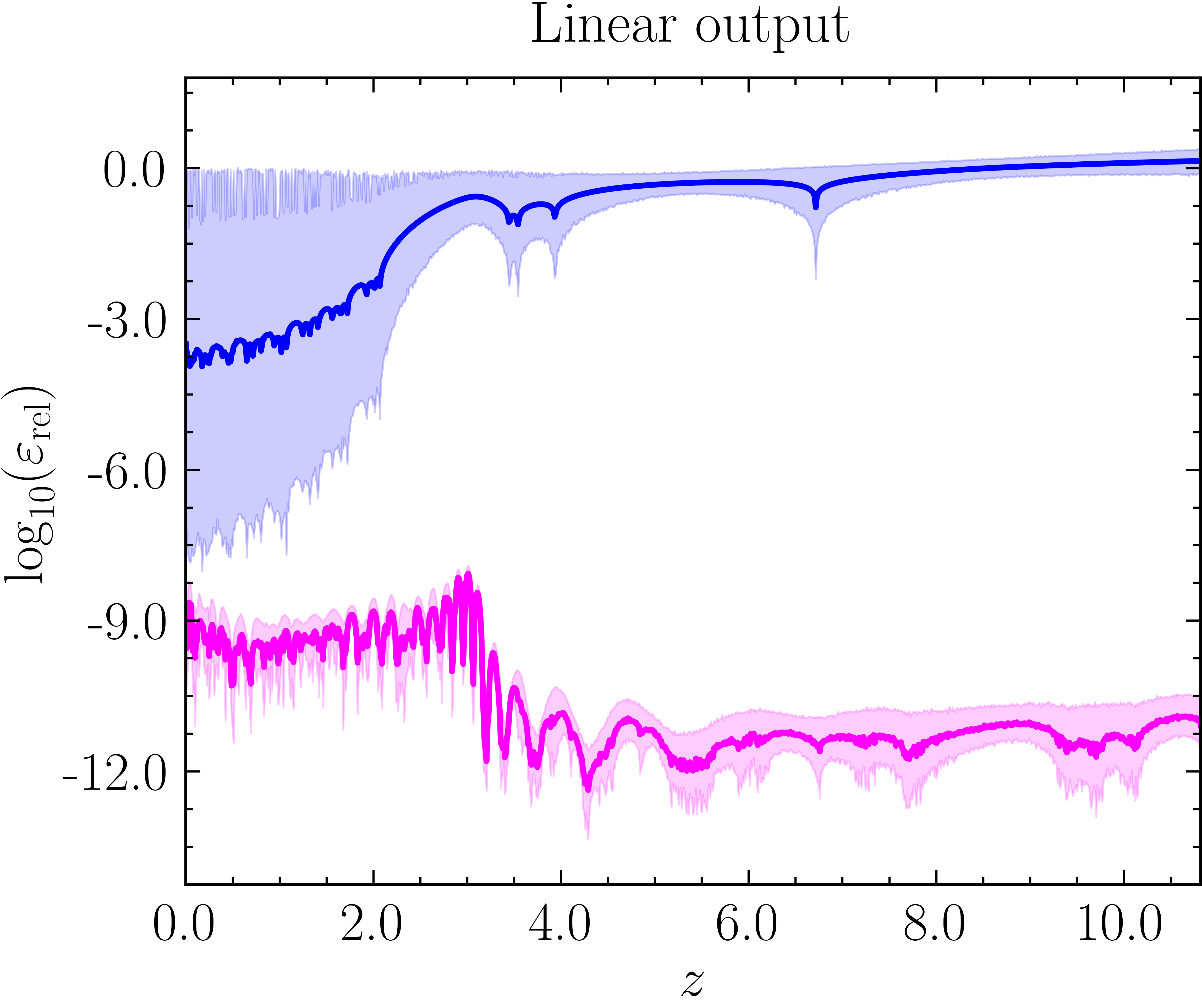} \hspace{0.1cm} \includegraphics[scale=0.29]{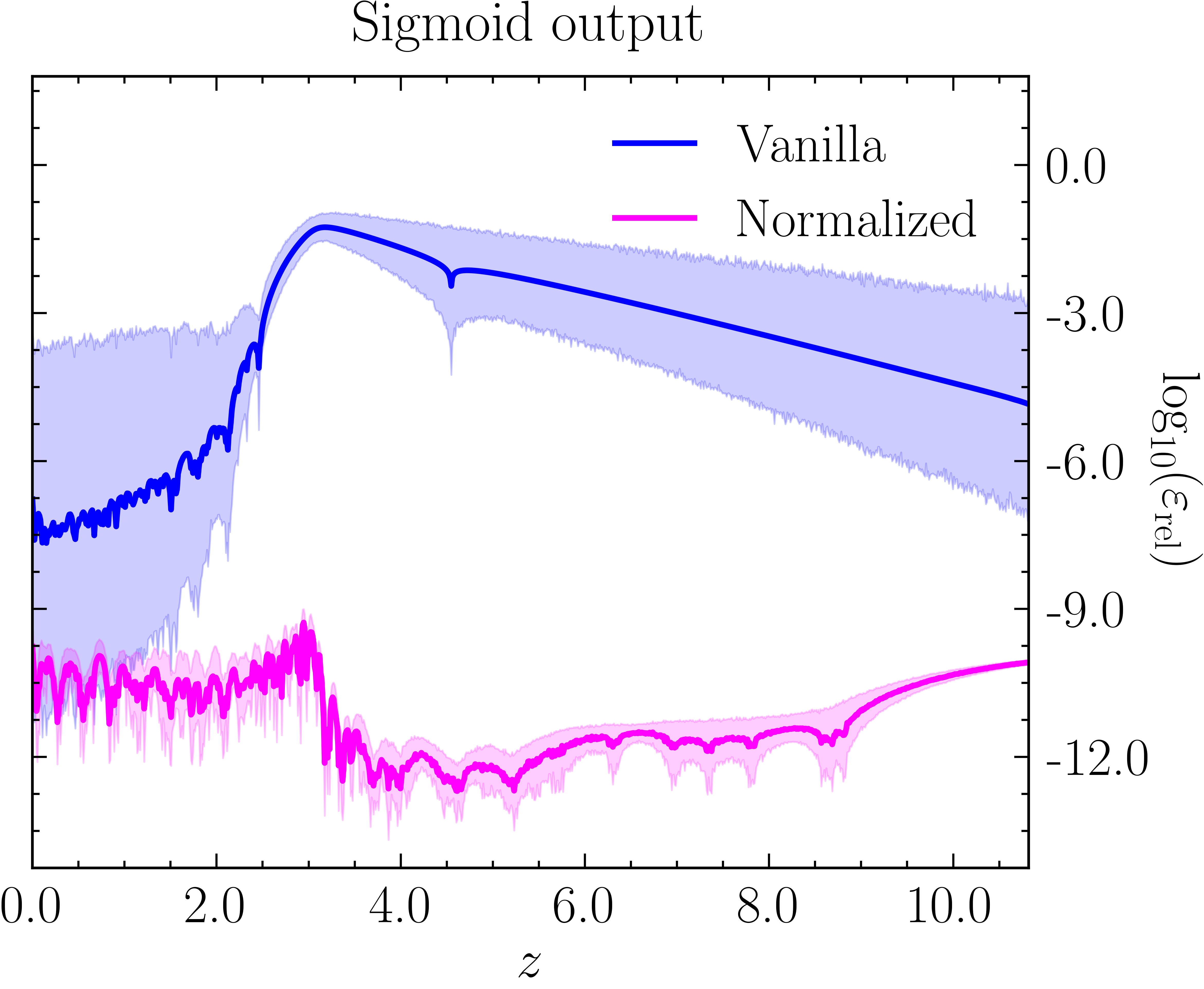}
    \caption{PINN $95\%$ confidence intervals for $\varepsilon_\text{rel}(z)$~\eqref{eq:epsilonMSEzk} with (magenta) and without (blue) Jacobian normalization, using linear (left) or negative sigmoid (right) output layers.}
    \label{fig:AblationTest}
\end{figure*}
In this section, we perform an ablation study by evaluating the performance of a PINN with Jacobian normalization relative to its vanilla (unnormalized) counterpart. The residual term in the loss function to be minimized by the PINN is defined as
\begin{equation}
    \mathcal{L}_{\text{r}} = \frac{1}{m} \sum_{k=1}^{m} \left(\lambda_{z_k} \ \mathcal{E}\left[z_k, W(z_k), \frac{d W}{dz}(z_k) ; C, z_t, \gamma\right] \right)^2 \; .
    \label{eq:Lbe}
\end{equation}
For the vanilla case $\lambda_{z_k} = 1$, whereas for Jacobian normalization $\lambda_{z_k} = 1/\sqrt{1 + J_y^2}$, with
\begin{align}
    J_y &= -\exp\!\left[C - \gamma\, \text{ReLU}(z_t - z) - z\right]
    \left[\exp(W) + \exp(2W_{\text{eq}} - W)\right] \; .
    \label{eq:freezeout_jacobian}
\end{align}
Additionally, the initial condition of the BE is imposed via
\begin{equation}
    \mathcal{L}_{\text{ic}} = \left[W(z_0)-W_0\right]^2 \; ,
    \label{eq:Lic}
\end{equation}
where $W_0$ is obtained from Eq.~\eqref{eq:Yeq} evaluated at $z_0 = 0$. 

For this ablation study, we make use of two PINN architectures that share the same structure -- four hidden layers with 40 units each and GELU activations -- but differ only in their output function: one employs a linear output, while the other uses a negative sigmoid\footnote{Since the normalized yield $W_n(z)$ lies in $[-1,0]$ (see Section~\ref{sec:Recast}), a negative sigmoid naturally enforces the required output range.}. The networks are trained with Adam optimizer for $2\times10^5$ epochs, starting with a learning rate of $\eta=0.001$ and decaying by $0.99$ every $1000$ epochs. We take $m=1000$, with the collocation points uniformly distributed over the interval $[0, \ln 5\times 10^4]$. Furthermore, we consider Standard cosmology ($\gamma = 0$) with $C = 29.2037$ chosen to reproduce the observed DM abundance (see Section~\ref{sec:inverse}). The training process is repeated $10$ times for both the vanilla and Jacobian normalized PINNs. At the end of each run we compute the square difference between the PINN and FEM as:
\begin{equation}
\varepsilon_\text{rel}(z_k) = \left[\frac{W_n(z_k)^{\text{PINN}} - W_n(z_k)^{\text{FEM}}}{W_n(z_k)^{\text{FEM}}} \right]^2 \, .
\label{eq:epsilonMSEzk}
\end{equation}
In Fig.~\ref{fig:AblationTest}, we show the $95\%$ confidence intervals for $\varepsilon_\text{rel}(z_k)$, with the left (right) panel corresponding to the PINN with a linear (negative sigmoid) output. The results for the normalized and vanilla cases are indicated by the magenta and blue shaded regions, respectively. Jacobian-normalized PINNs exhibit a marked improvement in precision compared to unnormalized networks. Quantitatively, for linear (sigmoid) outputs, the normalized case achieves $\varepsilon_\text{rel}\in [10^{-13}, 10^{-8}]$ $\left(\varepsilon_\text{rel}\in [10^{-14}, 10^{-9}]\right)$, whereas the vanilla PINN reaches much larger values, $\varepsilon_\text{rel}\in [10^{-8}, 1]$ $\left(\varepsilon_\text{rel}\in [10^{-11}, 10^{-2}]\right)$. The $95\%$ confidence intervals are also significantly narrower, indicating stable predictions across runs and suggesting that GD is robust under Jacobian normalization. Introducing a negative sigmoid output further enhances performance, particularly in the unnormalized case, and reduces the squared difference between PINN and FEM predictions by nearly two orders of magnitude for $z\lesssim 3$ at the freeze-out temperature~\eqref{eq:analyticalapprox}. 

%%%%%%%%%%%%%%%%%%%%%%%%%%%%%%%%%%%%%%%%%%%%%%%%%%%%%%%%%%%%%%%%%%%%%%%%%%%%%
\subsection{Jacobian normalization versus attention mechanisms}
\label{sec:JacobianVSattention}
%%%%%%%%%%%%%%%%%%%%%%%%%%%%%%%%%%%%%%%%%%%%%%%%%%%%%%%%%%%%%%%%%%%%%%%%%%%%%

%
    \begin{figure*}[t!]
        \centering
        \includegraphics[scale=0.29]{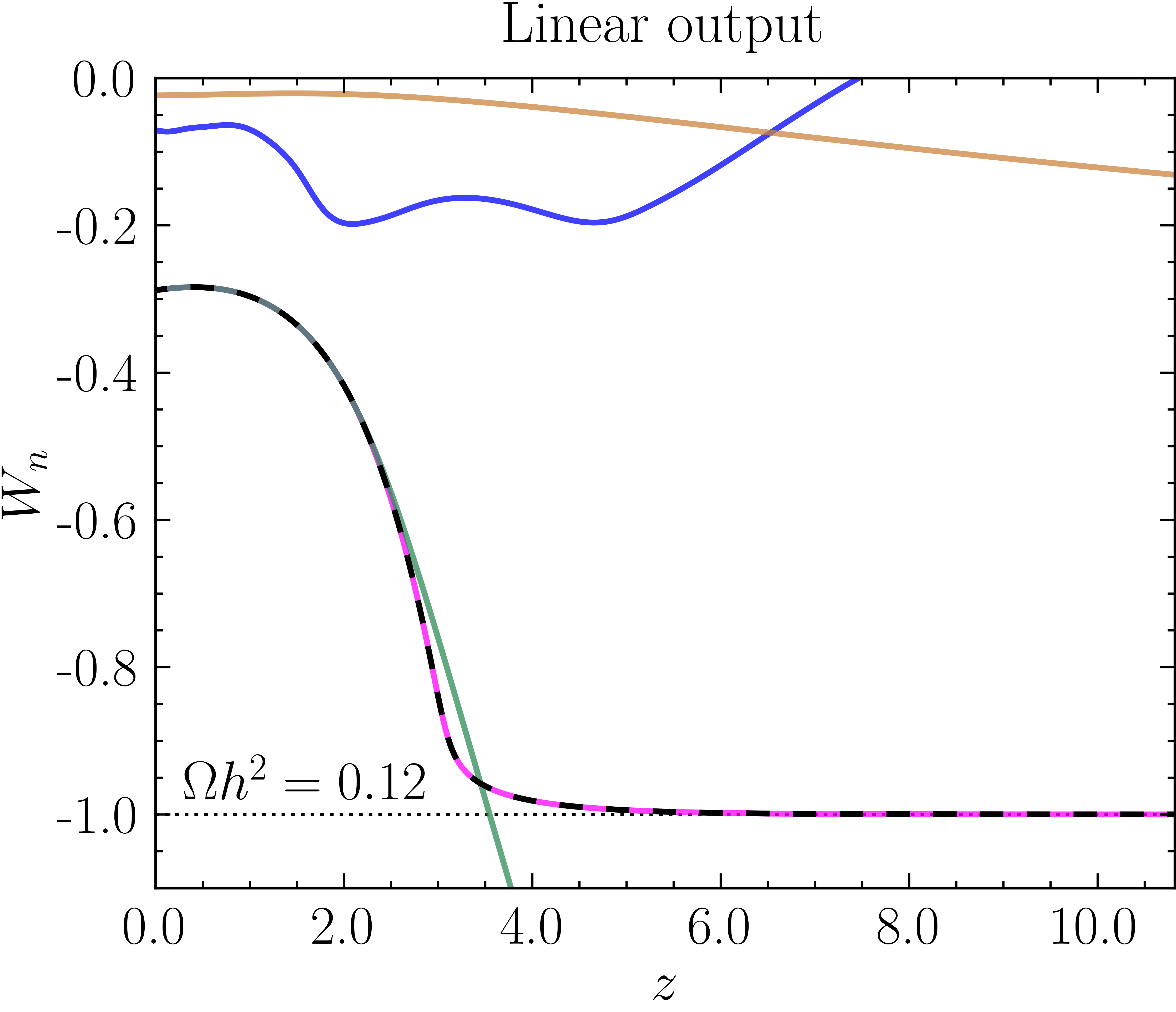} \hspace{0.1cm} \includegraphics[scale=0.29]{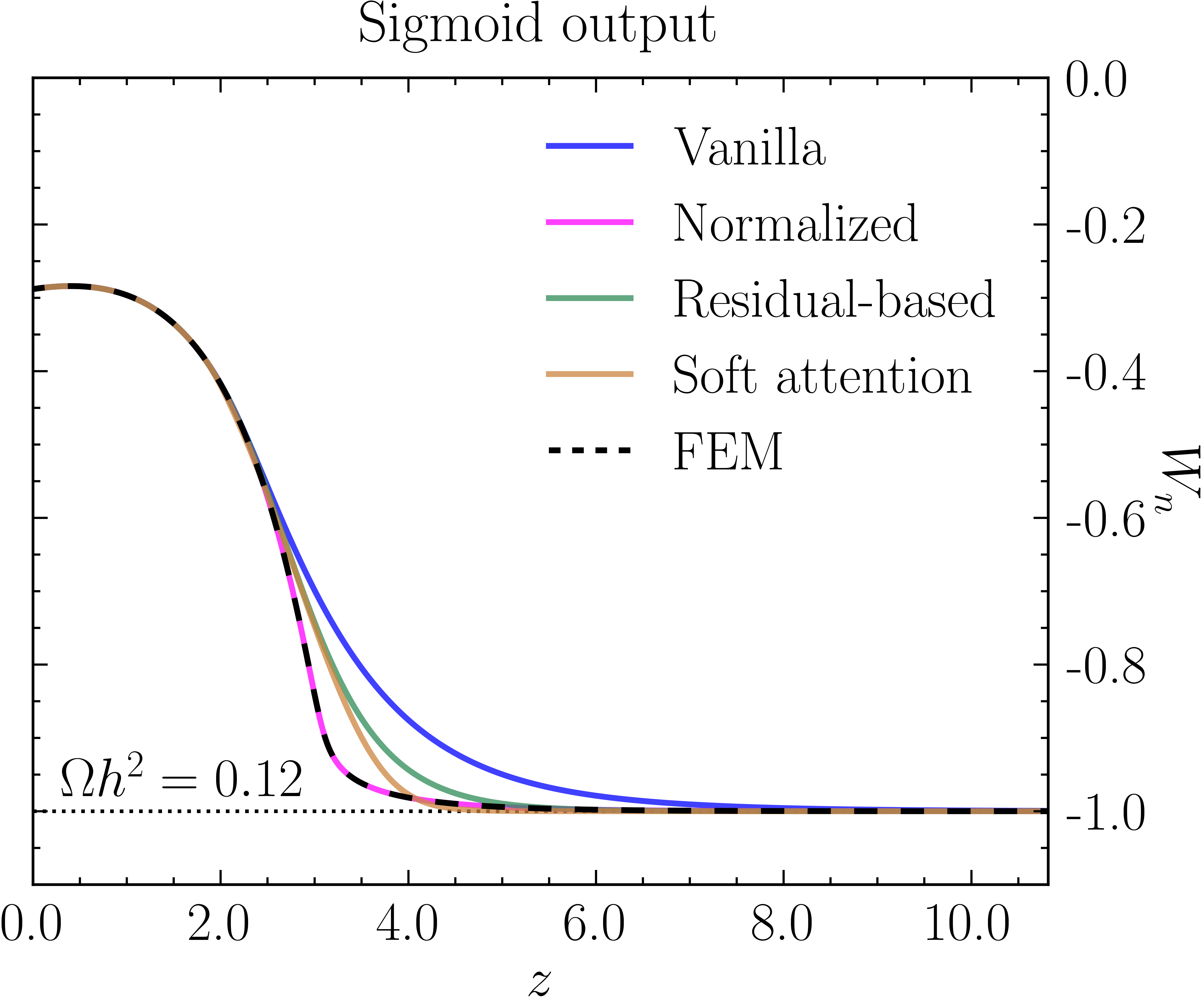} \\
        \vspace{+0.2cm} 
        \includegraphics[scale=0.29]{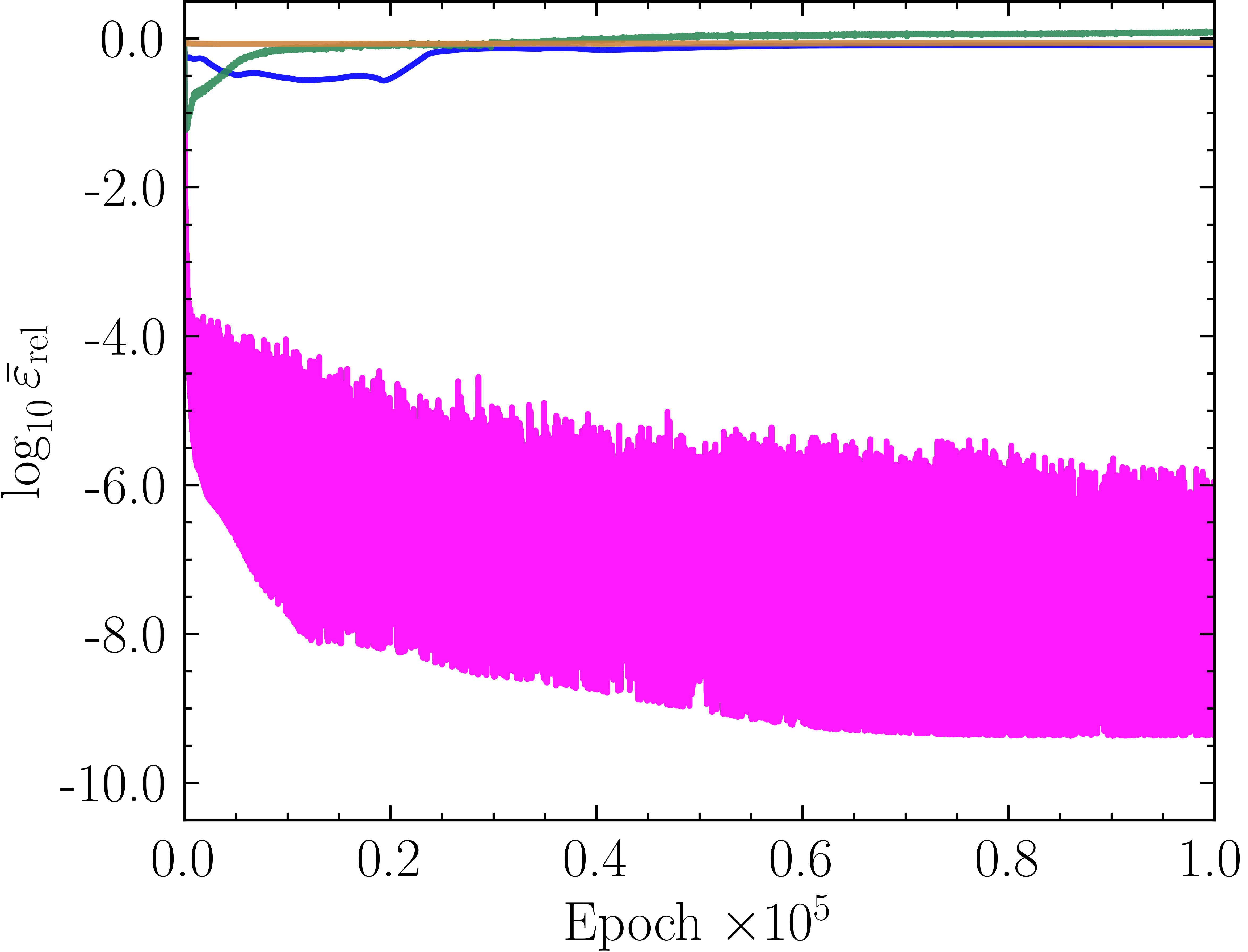} \hspace{-0.02cm} \includegraphics[scale=0.29]{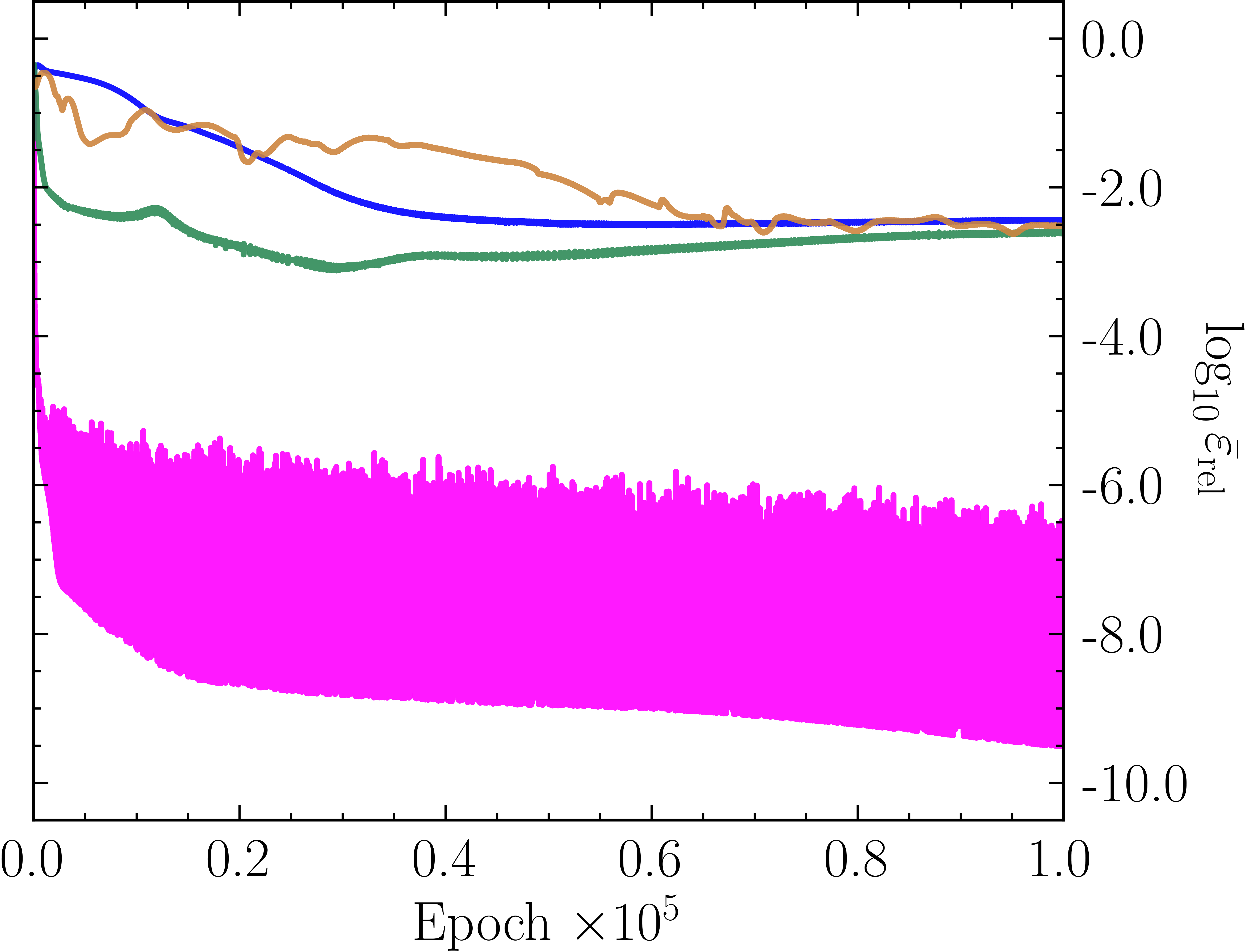} \\
        \vspace{+0.2cm} 
        \includegraphics[scale=0.29]{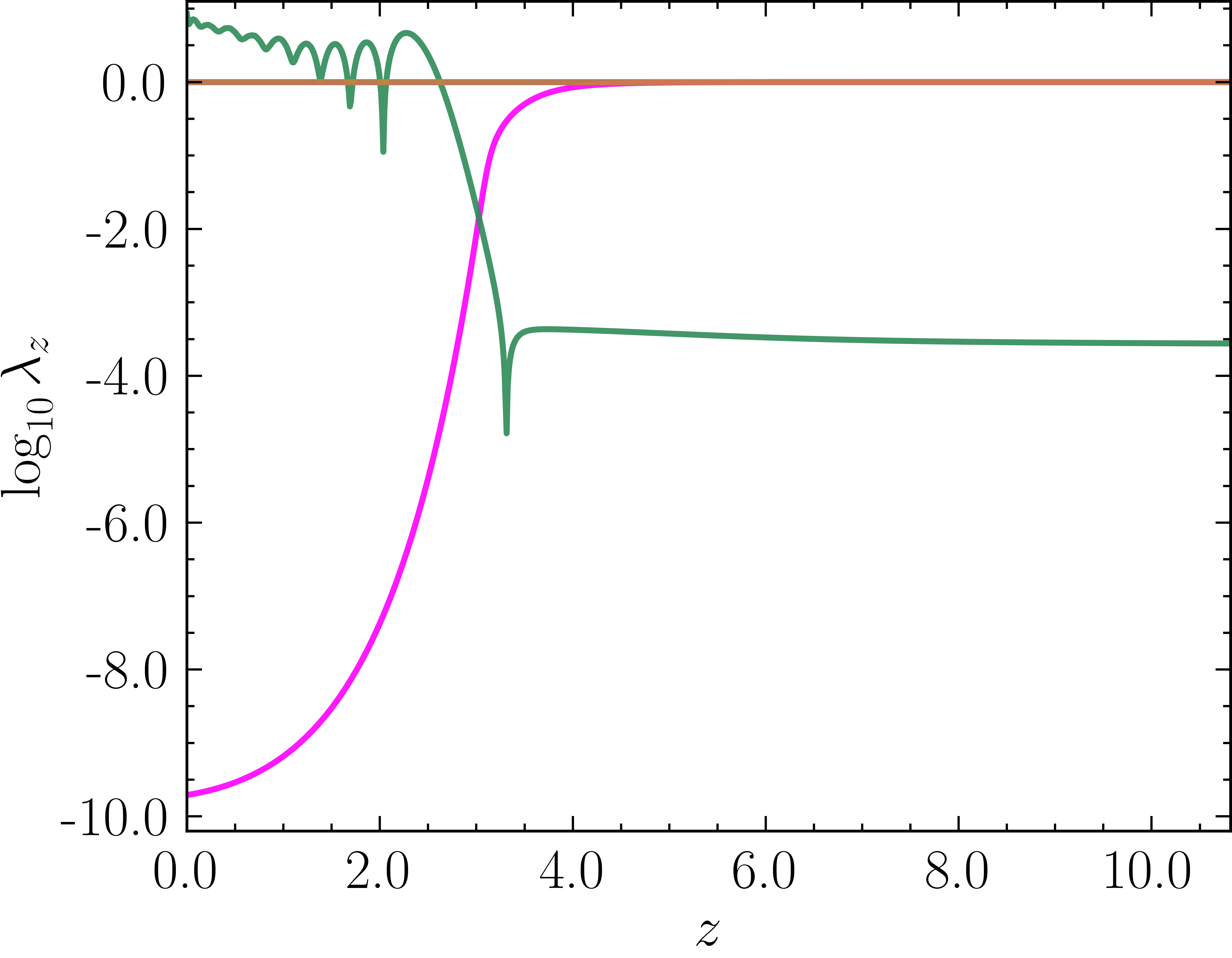} \hspace{0.1cm} \includegraphics[scale=0.29]{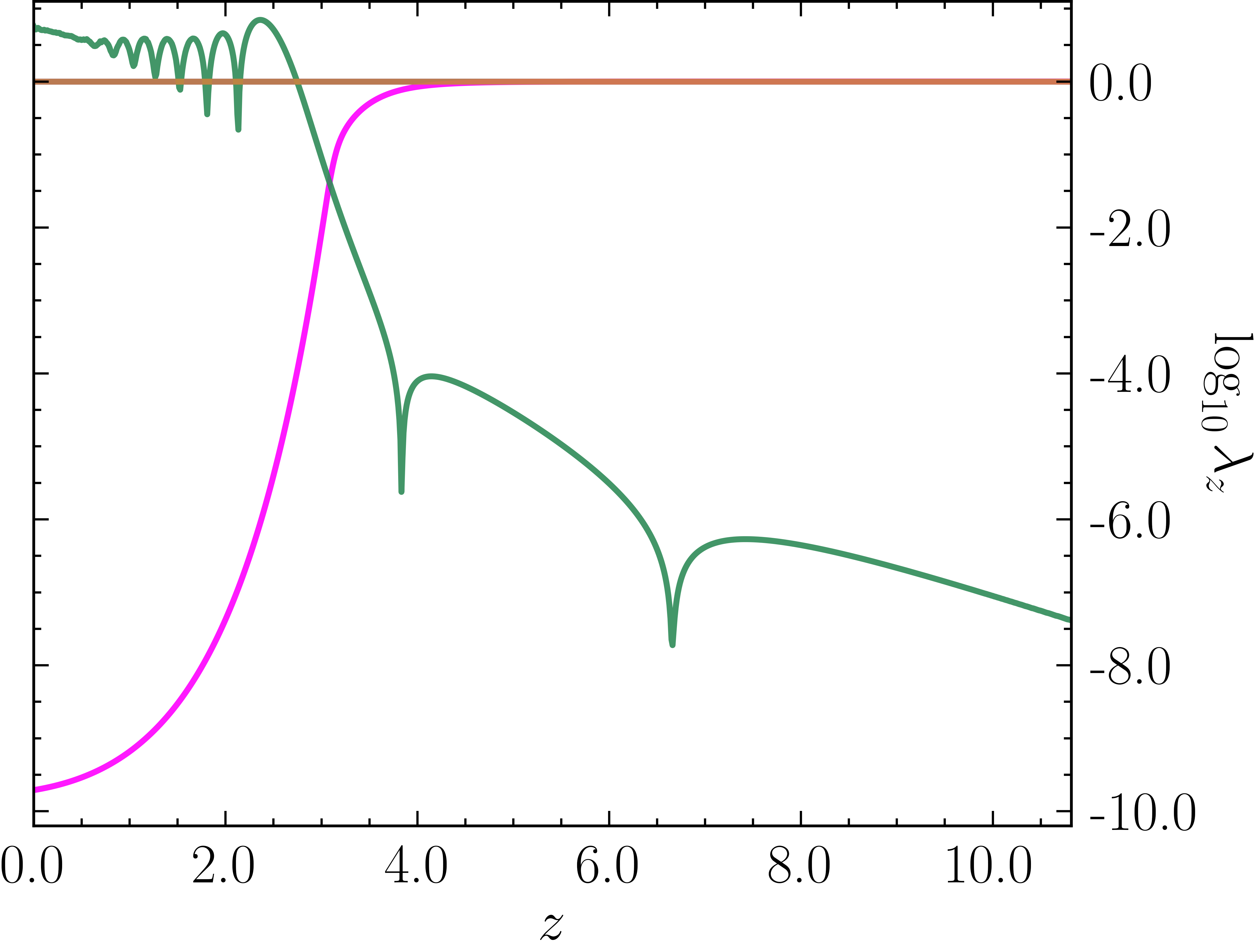}
        \caption{Solution of the freeze-out DM BE obtained with FEM (black dashed line) and various PINN approaches: vanilla (blue), Jacobian normalization (magenta), residual-based attention (green), and soft attention (brown). Left (Right): Results without (with) a negative sigmoid activation in the output layer. Top: Evolution of the WIMP DM particle yield. The horizontal dotted line satisfies the CDM abundance~\eqref{eq:Oh2exp}. Middle: Mean relative squared difference~\eqref{eq:epsilonMSE} between PINN and FEM during training. Bottom: Comparison of residual weights.}
    \label{fig:stiffness_comparison_results}
    \end{figure*}
In this section, we compare Jacobian normalization with two widely used approaches for addressing stiffness in PINNs, namely residual-based attention~\cite{anagnostopoulos2023residualbasedattentionconnectioninformation} and soft attention~\cite{McClenny_2023}. All three methods rely on the same underlying idea of assigning a weight to each collocation point, differing solely through their specific choice of weights $\lambda_{z_k}$ in Eq.~\eqref{eq:Lbe}. To implement self-adjusting weights for residual-based and soft attention, we follow Refs.~\cite{anagnostopoulos2023residualbasedattentionconnectioninformation} and~\cite{McClenny_2023}, respectively. In residual-based attention, $\lambda_{z_k} \in \mathbb{R}$ is a self-adaptive weight, while in soft attention they are given by $\lambda_{z_k} \rightarrow \sqrt{\text{sigmoid}(\lambda_{z_k})}$. Furthermore, soft attention replaces the initial condition loss in Eq.~\eqref{eq:Lic} with
\begin{equation}
    \mathcal{L}_{\text{ic}} = \lambda_\text{ic}\left[W(z_0)-W_0\right]^2 \; ,
    \label{eq:Lic2}
\end{equation}
where $\lambda_\text{ic} \in \mathbb{R}$ is a self-adaptive weight.

We use the two PINN architectures introduced in Section~\ref{sec:Ablation} and, we focus on the Standard cosmology case ($\gamma=0$) with $C=29.2037$. To track the training progress of each method, we compute the mean relative squared difference between the PINN predictions and the FEM results:
\begin{equation}
\overline{\varepsilon}_\text{rel} = \frac{1}{m}\sum_{k=1}^m \varepsilon_\text{rel}(z_k) \,,
\label{eq:epsilonMSE}
\end{equation}
where $\varepsilon_\text{rel}(z_k)$ is defined in Eq.~\eqref{eq:epsilonMSEzk}.

The results are shown in Fig.~\ref{fig:stiffness_comparison_results}, comparing PINN predictions for the vanilla (blue), Jacobian-normalized (magenta), residual-based attention (green), and soft-attention (brown) cases. Left (right) panels use linear (negative sigmoid) output activations. A few comments follow:

\begin{itemize}
    \item In the top panels, we show the WIMP DM particle yield $W_n(z)$. It is clear that only the Jacobian-normalized PINN (magenta) reproduces the FEM result (black dashed line), even without enforcing the negative-sigmoid output. Although the vanilla and attention-based PINNs improve significantly when the sigmoid output is used, they still fail to solve the BE accurately. These results make clear that only Jacobian normalization enables GD to converge reliably to the FEM solution.
    
    \item The middle panels display the relative error $\bar{\varepsilon}_{\mathrm{rel}}$~\eqref{eq:epsilonMSE} between the PINN solution and the benchmark FEM as a function of training epochs. Notice that, with Jacobian normalization (magenta), $\bar{\varepsilon}_{\mathrm{rel}}$ reaches its minimum value of $\mathcal{O}(10^{-8})$ very rapidly after only $\mathcal{O}(10^4)$ epochs. In contrast, for the other cases, $\bar{\varepsilon}_{\mathrm{rel}}$ remains at $\mathcal{O}(1)$ [$\mathcal{O}(10^{-2})$] at the end of training after $10^5$ epochs for linear [sigmoid] outputs. Once again, sigmoid output (right plots) yields improved performance for all cases, with the PINN accuracy, as quantified by $\bar{\varepsilon}_{\mathrm{rel}}$, enhanced by roughly one to two orders of magnitude.
    
    \item In the bottom panels we show the weights $\lambda_z$~\eqref{eq:Lbe} that minimize the loss for Jacobian-normalization and attention mechanisms. Recall that stiffness is associated with a large Jacobian [see Eq.~\eqref{eq:largejacobian}]. Since normalization assigns weights inversely proportional to the Jacobian, the magenta curve shows that stiffness decreases as $z$ increases. The attention-based schemes behave differently. For soft attention (brown), the weights remain fixed near 1, the maximum of the sigmoid mask function, effectively assigning no weight to the residual loss. In contrast, the behavior of $\lambda_z$ for residual-based attention (green) is opposite to the one of the normalized PINN, providing slightly worse results than soft attention.
    
\end{itemize}

%%%%%%%%%%%%%%%%%%%%%%%%%%%%%%%%%%%%%%%%%%%%%%%%%%%%%%%%%%%%%%%%%%%%%%%%%%%%%
\subsection{\textit{Inverse} problems: from the data to the possible physical models}
\label{sec:inverse}
%%%%%%%%%%%%%%%%%%%%%%%%%%%%%%%%%%%%%%%%%%%%%%%%%%%%%%%%%%%%%%%%%%%%%%%%%%%%%

%
    \begin{figure*}[t!]
        \centering
        \includegraphics[scale=0.155]{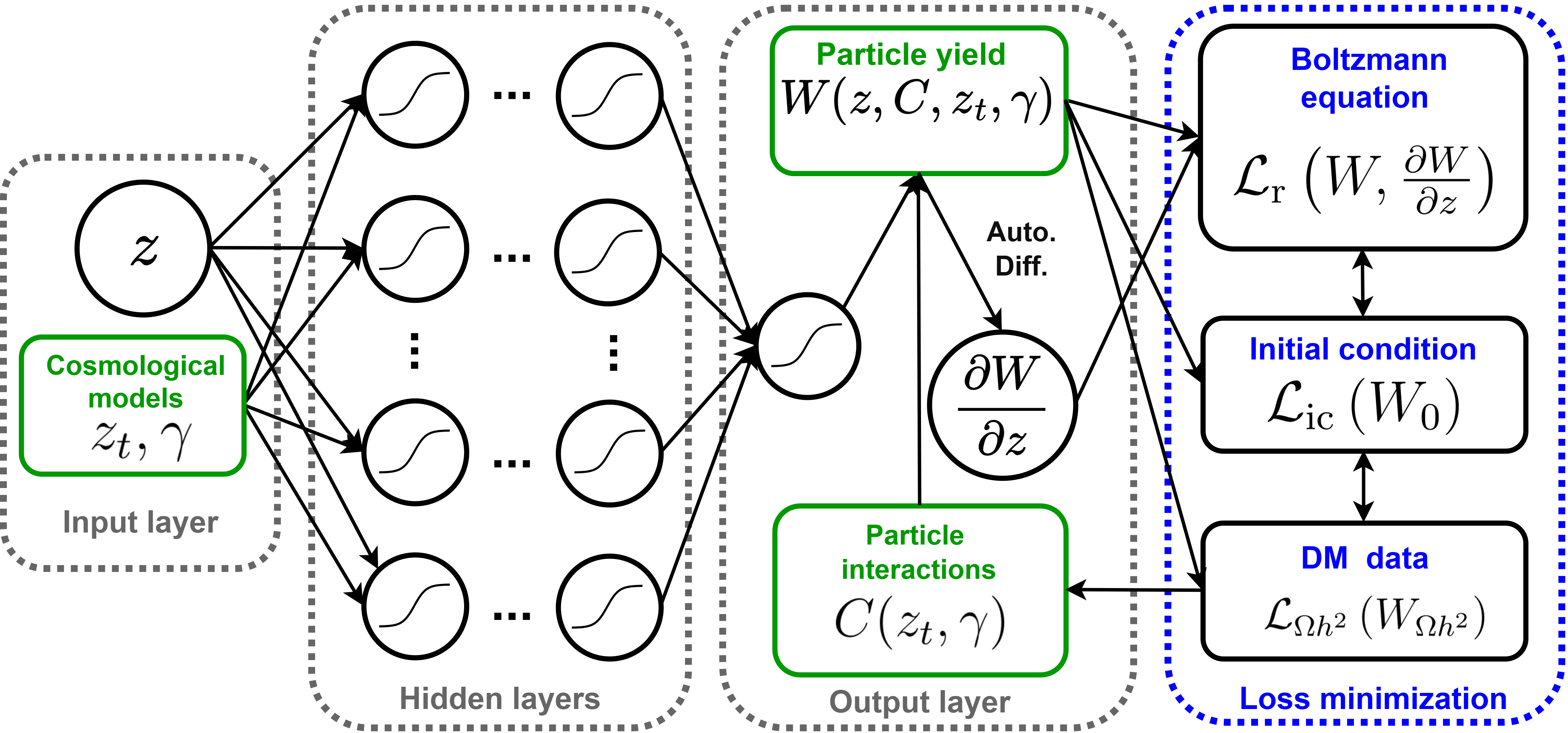}
        \caption{Diagrammatic representation of the \textit{inverse} PINN structure modeling the WIMP DM BE for a given cosmological scenario (see text for details).}
    \label{fig:inverse_PINN}
    \end{figure*}
We now highlight the performance of PINNs in addressing \textit{inverse} problems, emphasizing their utility in theoretical physics as tools for identifying models and/or parameter spaces. In contrast to the \textit{forward} problems discussed previously, where the cosmology and particle interactions of the BE are known and provided to the PINN before training, the \textit{inverse} formulation incorporates observational data -- here, the CDM abundance~\eqref{eq:Oh2exp} -- directly into the loss function, leaving certain theoretical parameters to be inferred by the network. The PINN architecture illustrated in Fig.~\ref{fig:inverse_PINN} schematically represents the \textit{inverse} problem: determining the particle interaction strength and DM yield given a cosmological model, such that the predicted CDM abundance matches the observed data. Specifically, we set \(z_t = \ln 500\) and \(\gamma = -2/3, 0, 2\), corresponding to the GB, Standard and RS cosmologies, respectively. The objective of the PINN is to infer the value of \(C(m,\langle\sigma v\rangle)\), now treated as a trainable parameter, by solving Eq.~\eqref{eq:freezeout2} while satisfying the experimental constraint~\eqref{eq:Oh2exp}.

Compared to the loss defined by Eqs.~\eqref{eq:Lbe} and~\eqref{eq:Lic}, the DM data point of Eq.~\eqref{eq:Oh2exp} enters the \textit{inverse} PINN through an additional term in the loss function (see Fig.~\ref{fig:inverse_PINN}):
\begin{equation}
    \mathcal{L}_{\Omega h^2} = \left[W(z_f) -W_{\Omega h^2}\right]^2 \; ,
    \label{eq:LOh2}
\end{equation}
measuring the squared difference between the PINN prediction at the last collocation point $W(z_f)$ and the observed DM relic density.

The set of collocation points used in the \textit{inverse} PINN is the same used in its \textit{forward} counterpart. Based on the analytical estimate given in Eq.~\eqref{eq:analyticalapprox} for Standard cosmology, we initialize $C$ following the normal prior
\begin{equation}
C\sim\mathcal{N}\left( 30.5,1.25 \right) \; ,
\label{eq:Cprior}
\end{equation}
such that the initial value has a $95 \%$ probability of lying within the interval $\left[28,33\right]$. Since $C$ is a trainable parameter, gradients propagate through the PINN to determine its update at the end of each epoch. Thus, we introduce an additional learning rate $\eta_C$ which acts on updating $C$. We set both learning rates starting at $\eta_C=\eta=0.002$, which decay at a rate of $0.99$ every 1000 epochs.

    \begin{figure*}[t!]
        \centering
        \includegraphics[scale=0.35]{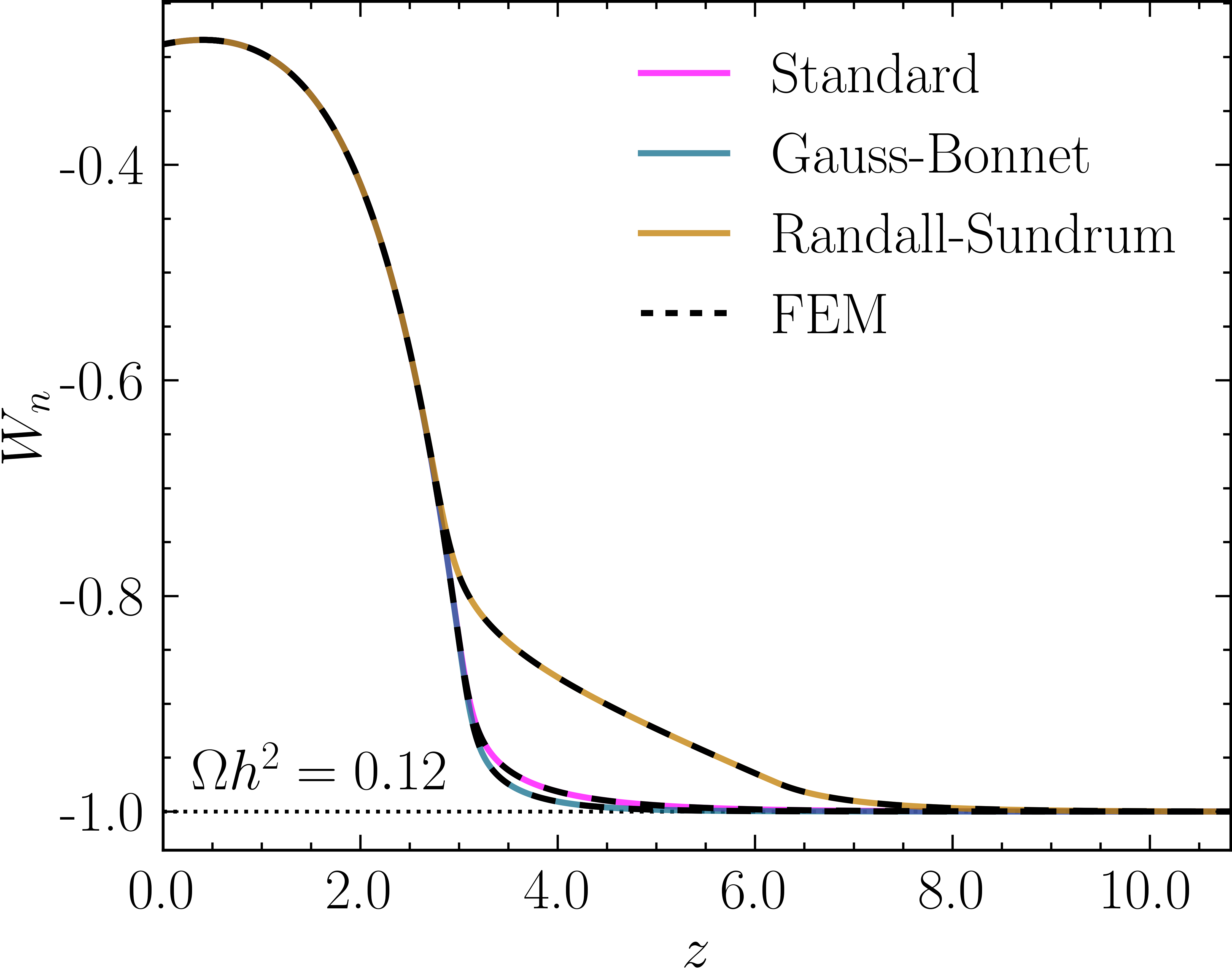}
        \caption{\textit{Inverse} PINN solution: particle yield evolution $W_n(z)$ for Standard (magenta), GB (green) and RS (brown) cosmology. We set $z_t = \ln 500$. Black-dashed curves were obtained via FEM. The horizontal dotted line corresponds to the $W_n(z)$ value reproducing the observed DM abundance~\eqref{eq:Oh2exp}.}
    \label{fig:inverse_PINN_results}
    \end{figure*}
\begin{table}[t!]
\centering
\renewcommand*{\arraystretch}{1.3}
\begin{tabular}{ccccc}  
\hline
$\gamma$ & $C_{\text{PINN}}$ & $\langle \sigma v \rangle$ $\left(\text{GeV}^{-2}\right)$ & $W(z_f)_{\text{FEM}}$ & $\Omega h^2_{\text{FEM}}$\\
\hline
$-2/3$ & $27.6359$ & $3.1174 \times 10^{-10}$ & $-26.1585$ & $0.1200$\\
$0$ & $29.2037$ & $1.4951 \times 10^{-9}$ & $-26.1585$ & $0.1200$\\
$2$ & $31.7022$ & $1.8187 \times 10^{-8}$ & $-26.1585$ & $0.1200$\\
\hline
\end{tabular}
\caption{\textit{Inverse} PINN solution: finding $C$ [see Eq.~\eqref{eq:Cdef}] for GB ($\gamma = -2/3$), Standard ($\gamma = 0$) and RS ($\gamma = 2$) cosmology --see Fig.~\ref{fig:inverse_PINN_results}. Last two columns cross-check the PINN results with FEM.}
\label{tab:inverse_predictions_gammaVsC}
\end{table}
In Fig.~\ref{fig:inverse_PINN_results}, we show the PINN solution to the \textit{inverse} problem of reconstructing the WIMP DM yield and determining the effective interaction strength $C$ for three cosmological scenarios: Standard (magenta), GB (green), and RS (brown). The corresponding $C$ values and derived cross sections, defined in Eq.~\eqref{eq:Cdef}, are summarized in Table~\ref{tab:inverse_predictions_gammaVsC}. To validate the PINN results, the predicted $C$ and $\gamma$ values were input into a FEM solver for the BE. The FEM output for the final yield $W(z_f)$ and the relic abundance provides a consistency check. Unlike PINNs, FEM solvers are not naturally suited for \textit{inverse} problems and typically require external optimization methods. The final column of the table shows that the relic abundances obtained using the PINN-derived parameters reproduce the experimental central value to within four decimal places, demonstrating the precision and reliability of the \textit{inverse} PINN approach. This is further confirmed in the figure since all three cosmological models yield relic abundances consistent with the observational constraint~\eqref{eq:Oh2exp}, lying well within the $1\sigma$ experimental uncertainty range (horizontal black dashed line). Notably, the results also show the influence of non-standard cosmological dynamics: in GB (RS) cosmology, characterized by a negative (positive) power-law index of $-2/3$ ($2$), a smaller (larger) interaction cross section is required compared to the Standard case to match the observed relic density. Overall, our analysis displays the complementarity between cosmology and particle physics in constraining viable DM models and how \textit{inverse} PINNs can serve as a valuable tool in theoretical physics model-building.

%%%%%%%%%%%%%%%%%%%%%%%%%%%%%%%%%%%%%%%%%%%%%%%%%%%%%%%%%%%%%%%%%%%%%%%%%%%%%
\section{Concluding remarks}
\label{sec:concl}
%%%%%%%%%%%%%%%%%%%%%%%%%%%%%%%%%%%%%%%%%%%%%%%%%%%%%%%%%%%%%%%%%%%%%%%%%%%%%

Stiff differential equations remain challenging for PINNs, often causing slow, unstable, or failed training. We introduced a \emph{Jacobian-based normalization of residuals} that mitigates stiffness by scaling the loss according to the local sensitivity of the governing equations. Analysis of the loss Hessian indicates improved gradient behavior during training. Unlike previous approaches, it requires no extra hyperparameters and works with analytical Jacobians or AD. Tests on various stiff ODEs show Jacobian normalization stabilizes training, accelerates convergence, and produces physically consistent solutions even in highly stiff regimes.

To showcase its relevance in theoretical physics, we applied our method to the freeze-out BE for WIMP DM, a paradigmatic stiff equation with no closed-form solution. Vanilla PINNs fail to converge, producing incomplete solutions, while our Jacobian-normalized formulation recovered the correct solution. Compared to conventional residual-based and soft attention methods, it achieved higher accuracy, stability, and faster convergence. Unlike FEMs, which target \textit{forward} problems, we showed \textit{inverse} PINNs can connect theory with experimental data. Using only the observed DM relic density~\eqref{eq:Oh2exp} as input, the network inferred the particle interaction strength in both Standard and alternative cosmologies, accurately reproducing the measured DM abundance.

Although \textit{inverse} PINNs have clear advantages over FEMs, limitations remain. Training is sensitive to hyperparameters, stiffness is not fully addressed for coupled ODEs and PDEs, and architectures often struggle to generalize across systems. Promising directions include extending \textit{Jacobian-based normalization} to coupled systems, integrating attention mechanisms and transfer learning, and applying curriculum learning for oscillatory systems. To make PINNs practical for DM phenomenology, further work must handle coupled systems, incorporate experimental constraints, and develop complete particle models with computed cross-sections.

In conclusion, \textit{Jacobian-based normalization} provides a simple way to enhance the robustness of
PINNs in stiff regimes, paving the way for more accurate and efficient modeling of dynamical systems
at the interface of particle physics and cosmology.

Code is available at~\url{https://github.com/MPedraBento/PINN-Jacobian-Normalization}.

%%%%%%%%%%%%%%%%%%%%%%%%%%%%%%%%%%%%%%%%%%%%%%%%%%%%%%%%%%%%%%%%%%%%%%%%%%%%%
\section*{Acknowledgments}
This research is supported by Fundação para a Ciência e a Tecnologia (FCT, Portugal) through the Projects UID/00777/2025 (https://doi.org/10.54499/UID/00777/2025) and CERN/FIS-PAR/0019/2021. The work of J.R.R. is supported by the PhD FCT grant 2025.00959.BD.
%%%%%%%%%%%%%%%%%%%%%%%%%%%%%%%%%%%%%%%%%%%%%%%%%%%%%%%%%%%%%%%%%%%%%%%%%%%%%

%%%%%%%%%%%%%%%%%%%%%%%%%%%%%%%%%%%%%%%%%%%%%%%%%%%%%%%%%%%%%%%%%%%%%%%%%%%%%
\bibliographystyle{utphys.bst}
\bibliography{bibliography.bib}
%%%%%%%%%%%%%%%%%%%%%%%%%%%%%%%%%%%%%%%%%%%%%%%%%%%%%%%%%%%%%%%%%%%%%%%%%%%%%

\end{document}